\documentclass[a4paper,11pt]{article}
\pdfoutput=1
\usepackage{jheppub}
\usepackage{amsmath}
\usepackage{amssymb}

\usepackage{physics}

\usepackage{graphicx}
\usepackage{dcolumn}
\usepackage{bm}
\usepackage{xcolor}
\usepackage{lineno}
\usepackage[normalem]{ulem}
\usepackage{cancel}
\usepackage{subcaption}

\newcommand{\orcid}[1]{\href{https://orcid.org/#1}{\includegraphics[width=8pt]{orcid.png}}}

\title{Testing non-standard neutrino interactions in (anti)-electron neutrino disappearance experiments}

\author{M. E. Chaves,}
\author{P. C. de Holanda,}
\author{and O. L. G. Peres}%
\affiliation{
Instituto de Física Gleb Wataghin - UNICAMP, 13083-859, Campinas SP, Brazil}

\emailAdd{mchaves@ifi.unicamp.br, ORCID:0000-0001-7396-081X}
\emailAdd{holanda@ifi.unicamp.br, ORCID: 0000-0001-9852-8900}
 \emailAdd{orlando@ifi.unicamp.br,ORCID:0000-0003-2104-8460}
 \abstract{ We search for scalar and tensor non-standard interactions using (anti)-electron neutrino disappearance in oscillation data. We found a slight preference for non-zero CP violation, coming from both tensor and scalar interactions. The preference for CP violation is lead by Daya Bay low-energy data with a significance that reaches $\sim1.7\sigma$ in the global analysis (and $\sim2.1\sigma$ when considering only medium baseline reactors data) compared to the standard neutrino oscillation scenario.
} 

\begin{document}

\newcommand{\fref}[1]{Fig.~\ref{fig:#1}} 
\newcommand{\eref}[1]{Eq.~\eqref{eq:#1}} 
\newcommand{\erefn}[1]{ (\ref{eq:#1})}
\newcommand{\erefs}[2]{Eqs.~(\ref{eq:#1}) - (\ref{eq:#2}) } 
\newcommand{\aref}[1]{Appendix~\ref{app:#1}}
\newcommand{\sref}[1]{Section~\ref{sec:#1}}
\newcommand{\cref}[1]{Chapter~\ref{ch:.#1}}
\newcommand{\tref}[1]{Table~\ref{tab:#1}}

\newcommand{\nn}{\nonumber \\}  
\newcommand{\nnl}{\nonumber \\}  
\newcommand{\nl}{& \nonumber \\ &}
\newcommand{\bnl}{\right .  \nonumber \\  \left .}
\newcommand{\dbnl}{\right .\right . & \nonumber \\ & \left .\left .}

\newcommand{\beq}{\begin{equation}} 
\newcommand{\eeq}{\end{equation}} 
\newcommand{\ba}{\begin{array}}  
\newcommand{\ea}{\end{array}} 
\newcommand{\bea}{\begin{eqnarray}}  
\newcommand{\eea}{\end{eqnarray} }  
\newcommand{\be}{\begin{eqnarray}}  
\newcommand{\ee}{\end{eqnarray} }  
\newcommand{\bal}{\begin{align}}
\newcommand{\eal}{\end{align}}   
\newcommand{\bi}{\begin{itemize}}  
\newcommand{\ei}{\end{itemize}}  
\newcommand{\ben}{\begin{enumerate}}  
\newcommand{\een}{\end{enumerate}}  
\newcommand{\bc}{\begin{center}}
\newcommand{\ec}{\end{center}} 
\newcommand{\bt}{\begin{table}}
\newcommand{\et}{\end{table}}  
\newcommand{\btb}{\begin{tabular}}
\newcommand{\etb}{\end{tabular}}  
\newcommand{\bvec}{\left ( \ba{c}}
\newcommand{\evec}{\ea \right )}

\newcommand{\cO}{{\mathcal O}} 
\newcommand{\co}{{\mathcal O}} 
\newcommand{\cL}{{\mathcal L}} 
\newcommand{\cl}{{\mathcal L}} 
\newcommand{\cM}{{\mathcal M}}

\newcommand{\const}{\mathrm{const}}

\newcommand{\mpl}{M_{\mathrm Pl}}

\def\mgut{\, M_{\rm GUT}}
\def\tgut{\, t_{\rm GUT}}
\def\mpl{\, M_{\rm Pl}}
\def\mkk{\, M_{\rm KK}}
\newcommand{\msusy}{M_{\rm soft}}

\newcommand{\dslash}[1]{#1 \! \! \! {\bf /}}
\newcommand{\ddslash}[1]{#1 \! \! \! \!  {\bf /}}

\def\ads{AdS$_5$\,}
\def\adse{AdS$_5$}
\def\intdk{\int {d^4 k \over (2 \pi)^4}} 

\def\ra{\rangle}
\def\la{\langle}  

\def\sgn{{\rm sgn}}
\def\pa{\partial}  
\newcommand{\dlr}{\overleftrightarrow{\partial}}
\newcommand{\Dlr}{\overleftrightarrow{D}}
\newcommand{\re}{{\mathrm{Re}} \,}
\newcommand{\im}{{\mathrm{Im}} \,}

\newcommand{\Ra}{\Rightarrow}
\newcommand{\lra}{\leftrightarrow}
\newcommand{\llra}{\longleftrightarrow}

\newcommand\simlt{\stackrel{<}{{}_\sim}}
\newcommand\simgt{\stackrel{>}{{}_\sim}}   
\newcommand{\zt}{$\mathbb Z_2$ }

\newcommand{\ha}{{\hat a}}
\newcommand{\hab}{{\hat b}}
\newcommand{\hac}{{\hat c}} 

\newcommand{\ti}{\tilde}  
\def\hc{{\rm h.c.}} 
\def\ov{\overline}  
  

\newcommand{\eps}{\epsilon}
\newcommand{\eS}{\epsilon_S}
\newcommand{\eT}{\epsilon_T}
\newcommand{\eP}{\epsilon_P}
\newcommand{\eL}{\epsilon_L}
\newcommand{\eR}{\epsilon_R}
\newcommand{\teps}{{\tilde{\epsilon}}}
\newcommand{\teS}{{\tilde{\epsilon}_S}}
\newcommand{\teT}{{\tilde{\epsilon}_T}}
\newcommand{\teP}{{\tilde{\epsilon}_P}}
\newcommand{\teL}{{\tilde{\epsilon}_L}}
\newcommand{\teR}{{\tilde{\epsilon}_R}}
\newcommand{\eLc}{{\epsilon_L^{(c)}}}
\newcommand{\eLv}{{\epsilon_L^{(v)}}}
\newcommand{\eSP}{\epsilon_{S,P}}
\newcommand{\teSP}{{\tilde{\epsilon}_{S,P}}}

\newcommand{\lz}{\lambda_z}
\newcommand{\dgz}{\delta g_{1,z}}
\newcommand{\dkg}{\delta \kappa_\gamma}

\def\cog{\color{OliveGreen}}
\def\cor{\color{Red}}
\def\copu{\color{purple}}
\def\coro{\color{RedOrange}}
\def\coma{\color{Maroon}}
\def\cob{\color{Blue}}
\def\cobr{\color{Brown}}
\def\cobl{\color{Black}}
\def\cost{\color{WildStrawberry}}

\newcommand{\tl}{{\tilde{\lambda}}}
\newcommand{\dll}{{\frac{\delta\lambda}{\lambda}}}

\maketitle


\section{Introduction}
\label{sec:introduction}

Since the establishment of the neutrino flavour oscillations \cite{Fukuda:1998mi,Ahmad:2002jz}, the effect was consistently probed by several experiments~\cite{kamland2013,DoubleChooz:2019qbj,PhysRevLett.121.241805,PhysRevLett.121.201801,T2K:2021xwb,NOvA:2019cyt,MINOS:2020llm,Hosaka:2005um,Fukuda:2002pe,Abe:2016nxk,Aharmim:2011vm,Agostini:2018uly,Kaether:2010ag,Abdurashitov:2009tn,Cleveland:1998nv}. In the present scenario, there are six measured parameters: $\theta_{13}$, $\theta_{12}$, $\theta_{23}$, $\Delta m^2_{21}$ and $\Delta m^2_{3i}$(i=1,2) in case of normal (inverted ) ordering~\cite{Esteban:2020cvm}. Remains unknown, the CP violation phase and the neutrino mass-ordering, which can be determined in the next few decades by Hyper-Kamiokande~\cite{Hyper-Kamiokande_2015}, DUNE~\cite{DUNE:2018tke,DUNE:2021tad}, and JUNO~\cite{JUNO_2016}. The neutrino oscillation experiments offer a new opportunity to search for effects of new physics coming from the neutrino sector, e.g, neutrino decays \cite{ND_1,ND_2,ND_3} and non-standard neutrino interactions (NSI) \cite{NC_NSI_1,NC_NSI_2,NC_NSI_3,NC_NSI_4,NC_NSI_5,NC_NSI_6,NC_NSI_7,NC_NSI_8,NC_NSI_9,NC_NSI_10,NC_NSI_11}. 

The NSIs can be classified according to its effects on the experiments: as \textit{interactions in the production and detection}  (mostly charged current) or \textit{interactions in the propagation} (mostly neutral current). Regarding neutrino interactions in the production and detection, it's frequent the search for left-handed interactions \cite{Langacker:1988up,Bergmann:1998ft,Grossman:1995wx,JOHNSON1998355,Johnson:1999ci,Gonzalez-Garcia:2001snt,Antusch_2006,Fernandez_Martinez_2007,Kopp_2008,Rodejohann_2010,Ohlsson:2012kf,Guzzo:2013tca,Agarwalla:2014bsa,EscrihuelaFerrandiz:2016pun,Rodejohann_2017,Falkowski:2019kfn,Ellis:2020ehi,Escrihuela:2021mud}, vector interactions~\cite{Khan:2017oxw,Khan:2013hva,Khan:2016uon,Khan:2021wzy}, scalar and tensor interactions~\cite{Khan:2019jvr,Falkowski:2019xoe,Du:2020dwr}. In this context, the interactions can have CP violating effects as discussed in Ref.~(\cite{Khan:2016uon,Khan:2021wzy}) for the vector interactions. We are going to work with the scalar and tensor interactions in the context of Standard Model Effective Field Theories~\cite{Buchmuller:1985jz,Grzadkowski:2010es} applied to neutrino phenomenology~\cite{Falkowski:2019xoe,Falkowski:2019kfn,Du:2020dwr,Falkowski:2021bkq,Falkowski_DUNE}.

We search for scalar and tensor interactions in the (anti)-neutrinos production (and detection) for reactors \cite{PhysRevLett.121.241805,PhysRevLett.121.201801,DoubleChooz:2019qbj,kamland2013} and solar \cite{Hosaka:2005um,Fukuda:2002pe,Abe:2016nxk,Aharmim:2011vm,Agostini:2018uly,Kaether:2010ag,Abdurashitov:2009tn,Cleveland:1998nv} experiments. The scalar and tensor interactions provide a new source of CP violation that is not present in the standard neutrino oscillation scenario for the (anti)-neutrino disappearance. The Lagrangian considered in this work assuming charged current interactions, is:
\begin{eqnarray}
\label{eq:EFT_lweft}
{\cal L}
& = &
- \,2\sqrt{2}G_F V_{ud} \big \{
  (\bar u  \gamma^\mu P_L d)  (\bar \ell_\alpha  \gamma_\mu P_L \nu_\alpha) + {1 \over 2 } [\epsilon_S]_{\alpha \beta} (\bar u d) (\bar \ell_\alpha P_L  \nu_\beta)
\nnl
&&
~~~+\,  {1 \over 4} [\epsilon_T]_{\alpha \beta} (\bar u  \sigma^{\mu \nu} P_L d)   (\bar \ell_\alpha  \sigma_{\mu \nu} P_L \nu_\beta)
+ \hc  \big \}~,
\label{eq:NSIL}
\end{eqnarray}
where the first term is the usual Standard Model Lagrangian, the second (third term) is the scalar (tensor) interaction Lagrangian proportional to the scalar (tensor) relative coupling $[\epsilon_S]_{\alpha \beta} ([\epsilon_T]_{\alpha \beta})$. We normalized the BSM terms by the Fermi coupling constant $G_F$,  and $V_{\rm ud}$ is a Cabibbo-Kobayashi-Maskawa~\cite{PhysRevLett.10.531,10.1143/PTP.49.652} (CKM) matrix element. The tensor structure is given by $\sigma^{\mu\nu}=i[\gamma^\mu,\gamma^\nu]/2$, and $P_{L,R}$ are the chirality projectors $(1\mp\gamma_5)/2$.

The article is organized as follows: in Sec. \ref{sec:formalism}, we introduce the formalism used in this work where we apply for the electron (anti)-neutrino disappearance. In Sec. \ref{sec:analysis}, we present the details of our analysis for reactor experiments as Daya Bay, RENO, and Double Chooz, as well as KamLand and for solar neutrino experiments. In Sec. \ref{sec:results}, we present our results and conclude in Sec. \ref{sec:conclusions}.

\section{BSM physics effects on neutrino oscillation}
\label{sec:formalism}
The effects of the BSM physics in the detection and production of the neutrino was computed following the References~\cite{Falkowski:2019xoe,Falkowski:2019kfn}. We first compute the rate for a $\nu_{\alpha}$  production followed by a $\nu_{\beta}$ detection in the Standard Model 
\begin{equation}
    R_{\alpha\beta}^{\rm SM}\propto \sum_{k,l} \int d\Pi_{P}d\Pi_D \left(\mathcal{M}_{\alpha k}^P\right)^{\rm SM}
   \left( \overline{\mathcal{M}}_{\alpha l}^P\right)^{\rm SM} \left(\mathcal{M}_{\beta k}^D \right)^{\rm SM}
   \left(\overline{\mathcal{M}}_{\beta l}^D\right)^{\rm SM}=\phi_\alpha^{\rm SM}\sigma_\beta^{\rm SM}
    \label{eq:total_ratesSM}
\end{equation}
where $\left(\mathcal{M}_{\alpha k}^{P,D}\right)^{\rm SM}$ are the amplitudes for production of a $\nu_{\alpha}$ neutrino and the detection of a $\nu_{\beta}$ neutrino in the Standard Model, $\phi_\alpha^{\rm SM}$ is the  $\nu_{\alpha}$ neutrino flux and $\sigma_\beta^{\rm SM}$ the detection cross-section for $\nu_{\beta}$ neutrino. 
Notice that there is no dependence on the distance between the source and the detection of the neutrino.
Likewise, the rates with BSM physics for a neutrino produced as $\nu_{\alpha}$ state, having travelled a distance $L$ and detected as  $\nu_{\beta}$ state is
\begin{equation}
    R_{\alpha\beta}\propto \sum_{k,l}e^{-i L \phi_{kl}}\int d\Pi_{P}d\Pi_D\mathcal{M}_{\alpha k}^P\overline{\mathcal{M}}_{\alpha l}^P\mathcal{M}_{\beta k}^D\overline{\mathcal{M}}_{\beta l}^D,
    \label{eq:total_rates-BSM}
\end{equation}
where $\mathcal{M}_{\alpha k}^{\rm P}(\mathcal{M}_{\beta k}^{\rm P})$ is the production, $\rm P$ (detection, $\rm D$),  amplitude for $\nu_{\alpha}(\nu_{\beta})$ for the BSM physics after the neutrino has travelled a distance $L$.  Here, $\phi_{kl}~\equiv~\Delta m^2_{kl}/2E_\nu$, where $\Delta m^2_{kl}\equiv m^2_k-m^2_l$ is the mass difference between the mass eigenstates $k$ and $l$, and $E_\nu$ is the neutrino energy. 

We can write the production amplitude of BSM physics as
\begin{equation}
   \mathcal{M}_{\alpha k}^{\rm P}=U_{\alpha i}{\cal M}_{\rm SM}^{\rm P}+\left[ \epsilon_{\rm X} U\right]_{\alpha i}{\cal M}_{X}^{\rm P}
    \label{amplitude-bsm}
\end{equation}
here, $U_{\alpha l}$ are components of the PMNS matrix~\cite{Pontecorvo:1967fh,1962PThPh..28..870M} that rotates the neutrino flavor fields into neutrino mass fields. Similarly, for the detection amplitude  $\mathcal{M}_{\alpha k}^{\rm D}$, replacing $U_{\alpha i}{\cal M}_{\rm SM}^{\rm P}\to U_{\alpha i}^*{\cal M}_{\rm SM}^{\rm D}$ and $\left[ \epsilon_{\rm X} U\right]_{\alpha i}{\cal M}_{X}^{\rm P}~ \to~\left[ \epsilon_{\rm X} U\right]_{\alpha i}^*{\cal M}_{X}^{\rm D}$. 

The relative change of the rate with BSM physics for a $\nu_{\alpha}$ being produced and $\nu_{\beta}$ being detected,  can be find replacing  Eq.~(\ref{amplitude-bsm}) into Eq.~(\ref{eq:total_rates-BSM}) and using Eq.~(\ref{eq:total_ratesSM}): 
\begin{flalign}
    \frac{R_{\alpha\beta}}{\phi_\alpha^{\text{SM}}\sigma_\beta^{\text{SM}}}=\sum_{k,l}{e^{-i\phi_{kl}L}}&\left[V_{\alpha}^{kl}(p_X)\right]
    \times \left[V^{kl}_{\beta}(d_X)\right]^*,
    \label{eq:rates}
\end{flalign}
where 
\begin{equation}
\begin{aligned}
    V_{\alpha}^{kl}(\rm p_X) = U^*_{\alpha k}U_{\alpha l}+p_{\rm XL}(\epsilon_X U)_{\alpha k}^*U_{\alpha l}+p_{\rm XL}^*U_{\alpha k}^*(\epsilon_X U)_{\alpha l}+p_{\rm XX}(\epsilon_X U)_{\alpha k}^*(\epsilon_X U)_{\alpha l},
\end{aligned}
\label{eq:vklpx}
\end{equation}
 and $p_{\rm XY}(d_{\rm XY})$ is the ratio of the production (detection) interference amplitude term $\mathcal{M}^X_{\alpha k}$ with $\mathcal{M}^Y_{\alpha k}$ over the SM amplitude. The equation \eqref{eq:vklpx} will be used in the place of the oscillation probability for the numerical simulations. The interaction amplitudes can be the standard model, scalar or tensor (X={ SM, S, T}) as given in Eq.~(\ref{eq:EFT_lweft}). We will call it {\it neutrino production and detection factors} and they are reaction dependent. They are 
 \begin{equation}
    \begin{aligned}
    p_{\rm XY} \equiv  \frac{\int d \Pi_{\rm P} A^P_{\rm X}\overline{A}^{\,
    \rm P}_{\rm Y}}{\int d\Pi_{\rm P}\left|A^{\rm P}_{\rm SM}\right|^2},& & d_{
    \rm XY} \equiv  \frac{\int d \Pi_{\rm D} A^{\, \rm D}_{
    \rm X}\overline{A}^{\, \rm D}_{\rm Y}}{\int d\Pi_{\rm D}\left|A^{
    \rm D}_{
    \rm SM}\right|^2},
    \end{aligned}
    \label{eq:std_nsi_ratio}
\end{equation}
For the expression for anti-neutrinos one should replace in Eq.~(\ref{eq:rates}):
\begin{equation}
    U\to U^* \quad \text{and} \quad  [\epsilon_X]_{\alpha \beta} \to [\epsilon_X]_{\alpha \beta}^*.
    \label{antineutrinos}
\end{equation}
The rate for NSI computed here, see Eq.~(\ref{eq:std_nsi_ratio}),  is the main difference between the usual assumption in the literature where we integrate out the fields and the parameters of NSI are independent of energy and universal for any production and detection process. Here the NSI contribution is dependent on the product 
\begin{eqnarray}
   [\epsilon_X]_{\alpha \beta}p_{\rm XY}
\end{eqnarray}
where the factor $[\epsilon_X]_{\alpha \beta}$ it is universal and energy independent, and the factor $p_{\rm XY}$ (or $d_{\rm XY}$) that it energy dependent and reaction dependent.
This formalism reduce to  literature if $p_{\rm XY}=1$ for any X and Y.

When turned off the BSM Lagrangian, $[\epsilon_X]_{\alpha \beta} \to 0$, the Eq.~(\ref{eq:rates}) return to expression of the usual neutrino conversion probability  $P(\nu_{\alpha}\to \nu_{\beta}$) for a initial neutrino $\nu_{\alpha}$ to a final neutrino $(\nu_{\beta})$. For non-zero BSM couplings, we can understand the right side of Eq.~(\ref{eq:rates}) as effective probabilities~\cite{Falkowski:2019kfn,Falkowski:2019xoe}. We caution the reader that this effective probabilities have non-unitary behaviour.

\begin{figure}[hbt]
    \centering
    \includegraphics{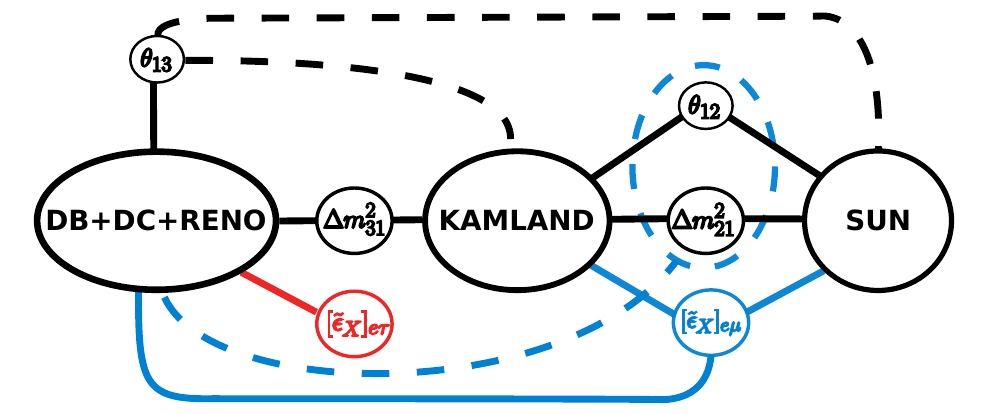}
    \caption{The solid (dashed) lines connect the relevant (less relevant) parameters to each experiment, Daya Bay (DB) \cite{PhysRevLett.121.241805}, RENO~\cite{PhysRevLett.121.201801}, Double Chooz (DC)~\cite{DoubleChooz:2019qbj}, KamLand  \cite{kamland2013} and solar experiments (SUN).}
    \label{fig:graph_param}
\end{figure}

In this work, we will assume a particular PMNS parametrization where the CP violation phase is enclosed in the $\theta_{23}$ rotation. Using this parametrization, is very convenient for studying the neutrino behaviour when we have (anti)-electron as booth, initial and final states. The alternative parametrization of PMNS matrix, will be defined by
\begin{equation}
U_{\rm PMNS}=U(\theta_{23},\delta) R (\theta_{13})R(\theta_{12})
     \label{bsm-pmns}
\end{equation}
where 
\begin{equation}
    \begin{aligned}
U(\theta_{23},\delta)=\begin{pmatrix}
    1&0&0\\
    0&c_{23}&s_{23}e^{i\delta}\\
    0&-s_{23}e^{-i\delta}&c_{23}
    \end{pmatrix},\;\;
&
R_{13}=\begin{pmatrix}
    c_{13}&0&s_{13}\\
    0&1&0\\
    -s_{13}&0&c_{13}
    \end{pmatrix},
&
    R_{12}=\begin{pmatrix}
    c_{12}&s_{12}&0\\
    -s_{12}&c_{12}&0\\
    0&0&1
    \end{pmatrix},
    \end{aligned}
    \label{pmns-coloma}
\end{equation}
where $c_{ij}\equiv \cos \theta_{ij}, s_{ij}\equiv \sin \theta_{ij}$ for i,j=1,2,3. This parametrization is also used in the reference~\cite{Coloma:2016}. The usual PMNS parametrization is $U_{\rm PMNS}=R(\theta_{23}) U (\theta_{13},\delta)R(\theta_{12})$ that we can get from Eq.~(\ref{pmns-coloma}) changing $U(\theta_{23},\delta)\to R(\theta_{23})$ and $R_{13}\to U(\theta_{23},\delta)$. The quantity $V_{\alpha}^{kl}(\rm p_X)$, defined in Eq.~(\ref{eq:vklpx}) will depend on the PMNS matrix $U$ and on the BSM coupling $[\epsilon_X]_{\alpha \beta}$ (as $\epsilon_X U$). Using the Eq.~(\ref{bsm-pmns}) to write down
\begin{equation}
(\epsilon_X U)\equiv [\epsilon_X] U= [\epsilon_X] 
U(\theta_{23},\delta) R (\theta_{13})R(\theta_{12})=
 [\tilde{\epsilon}_X]R (\theta_{13})R(\theta_{12}),
\label{eq:vklpx1}
\end{equation}
where we define the effective BSM parameter, 
\begin{equation}
[\tilde{\epsilon}_X]\equiv  [\epsilon_X] 
U(\theta_{23},\delta) ,
\label{eq:vklpx2}
\end{equation}
leading to the parameters to be the same as defined in \cite{Falkowski:2019xoe} and \cite{Khan:2013hva,Khan:2017oxw}
\begin{equation}
\begin{aligned}
 &[\tilde{\epsilon}_X]_{e\mu}=c_{23}[{\epsilon}_X]_{e\mu}-s_{23}[{\epsilon}_X]_{e\tau}e^{-i\delta},
 \nonumber \\
 &[\tilde{\epsilon}_X]_{e\tau}=c_{23}[{\epsilon}_X]_{e\mu}e^{i\delta}+s_{23}[{\epsilon}_X]_{e\tau},
\end{aligned}
    \label{eq:mynsi}
\end{equation}
the BSM effective coupling $(\epsilon_X U)$ can be computed for initial and final electron-neutrinos as 
\begin{eqnarray}
    (\tilde{\epsilon}_XO_{13}O_{12})_{e1}&=&-(s_{12}[\tilde{\epsilon}_X]_{e\mu}+c_{12}s_{13}[\tilde{\epsilon}_X]_{e\tau}),
\label{eq:nsinew1}
\\
    (\tilde{\epsilon}_XO_{13}O_{12})_{e2}&=&+(c_{12}[\tilde{\epsilon}_X]_{e\mu}-s_{12}s_{13}[\tilde{\epsilon}_X]_{e\tau}),
    \label{eq:nsinew2}
\\
(\tilde{\epsilon}_XO_{13}O_{12})_{e3}&=&c_{13}[\tilde{\epsilon}_X]_{e\tau}.
\label{eq:nsinew3}
\end{eqnarray}
Then, {\it the significant parameters for our analysis} are the mixing angles and mass differences $\theta_{12},\theta_{13}, \Delta m^2_{31},\Delta m^2_{21}$ and real and imaginary part of BSM effective coupling, $[\tilde{\epsilon}_X]$. We graphically displayed this information in Figure~(\ref{fig:graph_param})
for the experiments used in this analysis: Daya Bay, RENO, Double Chooz, and KamLand and the solar experiments. We connect them by solid (dashed) lines, the parameters more (less)  relevant for a given experiment, as will be discussed afterward.  We highlight an unusual sensitivity in reactors experiment  Daya Bay (DB), RENO, and Double Chooz (DC) to  $\Delta m_{21}^2$ oscillations coming from the CP violation effect of the $[\tilde{\epsilon}_X]_{e\mu}$ parameter as will be exhaustively discussed in this paper. 
 The advantage to use the specific PMNS parametrization (\ref{bsm-pmns}) is now evident due all information on the mixing angle $\theta_{23}$, CP phase $\delta$ and on the BSM coupling $[\epsilon_S]$ is encoded in the effective BSM coupling $[\tilde{\epsilon}_X]$. 

As can be seen from Eqs. \eqref{eq:nsinew1} \eqref{eq:nsinew2} and \eqref{eq:nsinew1}, $[\tilde{\epsilon}_X]_{\mu\tau}$ and the diagonals $[\tilde{\epsilon}_X]_{\mu\mu}$ and $[\tilde{\epsilon}_X]_{\tau\tau}$ does not appear in the final expression for the electron neutrino disappearance.
The $[\epsilon_X]_{ee}$ parameter can appear but will not be considered here, given the more substantial constraints imposed by inverse beta decay, see Ref.~\cite{Falkowski:2019xoe}. 

\subsection{Neutrino oscillations in reactors}
\label{sec:reactors}
For reactors, we have the anti-electron neutrinos ($\overline{\nu}_e$) in the process, then we should use Eq.(\eqref{antineutrinos}) into  Eq.\eqref{eq:rates} to get the anti-electron expression. We will rewrite the BSM rate in a more appealing structure separating non-oscillation, oscillations, and CP violation terms, for $\alpha=\beta=\overline{e}$,
\begin{equation}
\begin{aligned}
    \frac{R_{ee}}{\phi_e^{\rm SM}\sigma_e^{\rm SM}} = N^{\rm non-osc}-\sum_{k>l}N^{\rm osc}_{\rm kl}\sin^2{\left(\frac{ \phi_{\rm kl} L}{2}\right)}+\sum_{k>l}N^{\rm CP}_{\rm kl}\sin{\left({\phi_{\rm kl} L}\right)},
    \end{aligned}
    \label{eq:rates_as_prob}
\end{equation}
here, the $N^{\rm non-osc}$ is the term that causes the zero distance effect~\cite{Langacker:1988up,Antusch_2006,Fernandez_Martinez_2007,Kopp_2008,Rodejohann_2010,Ohlsson:2012kf,Guzzo:2013tca,Agarwalla:2014bsa,EscrihuelaFerrandiz:2016pun,Falkowski:2019kfn,Ellis:2020ehi}, where even for $L\to 0$ we will have non-zero effect of the BSM physics and it is a consequence of the non-unitary behaviour of BSM rate. For standard neutrino oscillations we have $N^{\rm non-osc}=1$. The other terms are the $N^{\rm osc}_{\rm kl}$ and $N^{\rm CP}_{\rm kl}$, the former is the amplitude of oscillation and the latter is the amplitudes for the CP violation terms generated by the interference of the standard model amplitudes with the new interactions. 
The full expressions for the amplitudes, for the $\alpha=\beta=e$ case can be found in the Appendix \ref{sec:analytical}. 

\subsubsection{Reactors  neutrino production and detection factors}
\label{proddectmbr}

As mentioned before, in this formalism are present the production and detection factors that should be computed for each neutrino reaction. The BSM physics effects can be factorized in a reaction dependent factor $p_{\rm XL}$ and $d_{\rm XL}$, as given in Eq.~(\ref{eq:std_nsi_ratio}) and a universal factor,
$[\tilde{\epsilon}_X]$ as given Eq.~(\ref{eq:vklpx2}).

In reactors, we assume that anti-neutrinos are produced mainly from Gamow-Teller decays and the detection comes from inverse beta process (Gamow-Teller and Fermi): 
\begin{table}[hbt]
\centering
\begin{tabular}{|l|cc|l|cc|}
 & scalar & tensor & & scalar & tensor \\ \hline 
$p_{\rm XL}$  & $0$ & $-\dfrac{g_T}{g_A}\dfrac{m_e}{f(E_\nu)}$  & $p_{\rm XX}$  & $\dfrac{g^2_S}{3g_A^2}$ & $\dfrac{g^2_T}{g_A^2}$
\\&&&&&\\$d_{\rm XL}$  &  $\dfrac{g_Sg_V}{g_V^2+3g_A^2}\dfrac{m_e}{E_e}$ & $\dfrac{3g_Ag_T}{g_V^2+3g_A^2}\dfrac{m_e}{E_e}$ &
$d_{XX}$  & $\dfrac{g_S^2}{g_V^2+3g_A^2}$ & $\dfrac{3g_T^2}{g_V^2+3g_A^2}$
\end{tabular}
\caption{Production and detection factors, Eq. \eqref{eq:std_nsi_ratio} for reactors neutrinos. Here $E_e$ is the positron energy, $m_e$ is the electron mass and $f(E_\nu)$ take in to account the nuclear fuel~\cite{Falkowski:2019xoe} and sub-products in the reactor model, and the $g_A, g_V, g_T$ and $g_S$ were extracted from Refs.~\cite{Gonzalez-Alonso:2018omy,Bhattacharya:2016zcn,Gonzalez-Alonso:2013ura}. For low-energy, the positron energy can related to the neutrino electron by the relation $ E_e = E_\nu -\Delta_{\rm fi}$, where $\Delta_{\rm fi} = m_{N_f}-m_{N_i}$ is the mass difference between the final and initial nucleon.}
\label{tab:factorsractors}
\end{table}

\begin{equation}
    \begin{aligned}
        \beta^-:\;&^A_ZN_i \rightarrow ^A_{Z+1}N_f+e^-+\overline{\nu}_e, &\text{(production)}\\
        \text{inv. }\beta^+:\;&^A_ZN_i+\overline{\nu}_e\rightarrow e^++  ^A_{Z-1}N_f,&\text{(detection)}
    \end{aligned}
\label{eq:reactions_react}
\end{equation}
In our calculations, we use the Lagrangian of Eqs. \eqref{eq:NSIL} and compute the 
factors $p_{\rm XL}$ and $d_{\rm XL}$ given in Eq.~(\ref{eq:std_nsi_ratio}) for scalar and tensor interactions, where we neglect nuclear recoil effects. The results are in Table~(\ref{tab:factorsractors}, are {\it energy dependent and this is a distinct behaviour of this formalism}. The $p_{\rm XL}$ and $d_{\rm XL}$ coefficients are functions of the neutrino energy. 

Each of neutrino experiments that we are going to test have a different sensitivity to the two mass difference present in 3$\nu$ oscillation scenario, $ \Delta m^2_{31}$ and $ \Delta m^2_{21}$. In Daya Bay (DB) \cite{PhysRevLett.121.241805}, RENO~\cite{PhysRevLett.121.201801} and Double Chooz (DC)~\cite{DoubleChooz:2019qbj} experiments that are medium baseline reactor (MBR) experiments with  $\frac{L}{E_{\nu}} \sim \frac{1\rm km}{3\, \rm MeV}$, the product $\phi_{kl}L$ that appear in Eq.~\eqref{eq:total_rates-BSM}  can be written as 
\begin{equation}
 \phi_{31} L=4.3\, \left({ \Delta m^2_{31} \over  2.4\times 10^{-3}{\rm eV}^2}\right)\left( {\rm L \over 1000{\rm m}}\right)\left(\frac{3 {\rm MeV}}{\rm E}\right)\quad \quad \phi_{21}L\ll 10^{-2}\left(\frac{\rm L}{1000{\rm m}}\right)
 \label{phase31}
\end{equation}
Given that, the contribution to the the oscillation due the $\sin^2 (\phi_{kl}L)$ function comes only from the $\Delta m^2_{31}$ contribution. When we have BSM physics, for the medium baseline, the $N^{\rm osc}_{kl}$ term will be proportional to $(\tilde{\epsilon}_XO_{13}O_{12})_{e3}$ given in Eq.~(\ref{eq:nsinew3}), thereafter, proportional to $[\tilde{\epsilon}_X]_{e\tau}$:
\begin{equation}
    N_{\rm atm}^{\rm osc}=s^2_{2\theta_{13}}+2s_{2\theta_{13}}c_{2\theta_{13}}\Re \{[\tilde{\epsilon}_X]_{e\tau}\}({\rm d_{\rm XL}}+{\rm p_{\rm XL}})+\mathcal{O}([\tilde{\epsilon}_X]_{e\alpha}^2)
    \label{atmoscillations}
\end{equation}
where we define $s_{2\theta_{ij}}\equiv \sin\left(2\theta_{ij} \right),c_{2\theta_{ij}}\equiv \cos\left(2\theta_{ij}\right)$ and $\Re A (\Im A)$ is the real and imaginary part of variable A.
This result implies that these experiments can only test $[\tilde{\epsilon}_X]_{e\tau}$, but not the $[\tilde{\epsilon}_X]_{e\mu}$. 
For KamLand~\cite{kamland2013}, the more relevant oscillation scale is the $\phi_{21} L=5.8\,\left(\frac{\Delta m^2_{21}}{7.5\times 10^{-5}{\rm eV}^2}\right)\left( \frac{\rm L}{180{\rm km}}\right)\left(
\frac{3 {\rm MeV}}{\rm E}\right)$. The $N_{\rm sun}^{\rm osc}$ contribution for this experiment is
\begin{equation}
\begin{aligned}
    N_{\rm sun}^{\rm osc}=&
    c_{13}^3s_{2\theta_{12}}^2\left(
    c_{13}+2\left(t_{2\theta_{12}}\Re \{[\tilde{\epsilon}_X]_{e\mu}\}+s_{13}\Re \{[\tilde{\epsilon}_X]_{e\tau}\}\right)({\rm d_{\rm XL}}+{\rm p_{\rm XL}})+\mathcal{O}([\tilde{\epsilon}_X]_{e \alpha}^2)\right),
\end{aligned}
\end{equation}
where  we define $ t_{2\theta_{12}} \equiv \tan \left( 2\theta_{12}\right)$. Here,
the $[\tilde{\epsilon}_X]_{e\tau}$ contribution is suppressed by $s_{13}$ and for $[\tilde{\epsilon}_X]_{e\mu}$ is not suppressed, undergoing to a small effects of $[\tilde{\epsilon}_X]_{e\tau}$ in Kamland compared with $[\tilde{\epsilon}_X]_{e\tau}$.
In short, the medium baselines, Daya Bay (DB) \cite{PhysRevLett.121.241805}, RENO~\cite{PhysRevLett.121.201801} and Double Chooz (DC)~\cite{DoubleChooz:2019qbj} experiments are more sensitive to $[\tilde{\epsilon}_X]_{e\tau}$ and KamLand~\cite{kamland2013} is more sensitive to $[\tilde{\epsilon}_X]_{e\mu}$. 

In addition to the oscillation terms $N_{kl}^\text{osc}$, in Eq. \eqref{eq:rates_as_prob} there is CP violating term $\left(N^{\rm CP}_{kl}\right)$. 
In the standard case of oscillation, induced by PMNS mixing matrix, the CP violating amplitude is $\left(N^{\rm CP}_{kl}\right)^{\rm S.O.} \to \Im \left(J^{\alpha\beta}_{kl}\right)$ where $\left(J^{\alpha\beta}_{kl}\right)\equiv U_{\alpha l}^*U_{\beta l}U_{\alpha k} U_{\beta k}^*$ is the Jarskolg invariant~\cite{PhysRevLett.55.1039}. In our setup with $\alpha=\beta=e$, we should have $\left(N^{\rm CP}_{kl}\right)^{\rm S.O.}\to 0$, that is, we have {\it no CP violation for electron neutrino disappearance experiments in PMNS scenario}, but in BSM scenario this term is non-zero and comes from the interference between the standard model amplitude  with the BSM amplitude. 
For the $[\tilde{\epsilon}_X]_{e\mu}([\tilde{\epsilon}_X]_{e\tau})$ parameter the expression for the CP violating term is 
\begin{align}
   N^{\rm CP}_{\rm Sun} =& \left[c_{13}^2(d_{\rm XL}-p_{\rm XL})+\left|[\tilde{\epsilon}_X]_{e\mu}\right|^2(d_{\rm XX}p_{\rm XL}-d_{
   \rm XL}p_{\rm XX})\right]\Im{[\tilde{\epsilon}_X]_{e\mu}}c_{13}{s_{2\theta_{12}}},\label{eq:CP1}\\
    N^{\rm CP}_{\rm atm} =& \left[(d_{\rm XL}-p_{\rm XL})+s_{2\theta_{13}}\left|[\tilde{\epsilon}_X]_{e\tau}\right|^2(d_{\rm XX}p_{\rm XL}-d_{\rm XL}p_{\rm XX})\right]\Im{[\tilde{\epsilon}_X]_{e\tau}}s_{2\theta_{13}}.
  \label{eq:CP2}
\end{align}
Here, is explicit that the source of CP violation is the imaginary part of $[\widetilde{\epsilon}_X]_{e\beta}$.  Another thing that appear is that the CP violation terms are always proportional by the difference $(d_{XL}-p_{XL})$ or the difference  $( d_{XX}p_{XL} - p_{XX}d_{XL})$, these two differences are asymmetry between the production and detection process that mimic a T violation in production and detection. 

In principle, experiments as Daya Bay (DB) \cite{PhysRevLett.121.241805}, RENO~\cite{PhysRevLett.121.201801} and Double Chooz (DC)~\cite{DoubleChooz:2019qbj} are not sensitive to the solar scales terms, $\Delta m^2_{21}$. This is because of Eq.~(\ref{phase31}), the effects on the usual neutrino oscillation from the $\Delta m^2_{21}$ contribution is $(\phi_{21}L)^2$ that it too small. 
However, for CP violating terms, the amplitude $  N^{\rm CP}_{kl}$ it is multiplied instead by $\sin \left(\phi_{\rm kl} \right)$. Due this, the CP violating term can be at the same order of the usual oscillation term from $ \Delta m^2_{31}$, this happens when
\begin{equation}
    \Im \left\{[\tilde{\epsilon}_X]_{e\mu}\right\}s_{2\theta_{12}}\sin(\Delta_{\rm 21})\sim s^2_{2\theta_{13}}\sin^2(\Delta_{\rm 31}).
    \label{eq:CP_new}
\end{equation}
This may cause medium baseline reactors sensitive to CP-violation not only from the atmospheric scales but also from the {\it solar scale}. This is shown graphically in Fig.~(\ref{fig:graph_param}), where the medium baseline neutrino reactors experiments, Daya Bay (DB) \cite{PhysRevLett.121.241805}, RENO~\cite{PhysRevLett.121.201801} and Double Chooz (DC)~\cite{DoubleChooz:2019qbj}  are sensitive to $ \Delta m^2_{21}$.

From Eqs. \eqref{eq:CP1}, \eqref{eq:CP1} and from the discussions above, we conclude that for CP violating term, the KamLand~\cite{kamland2013} will more sensitive to $[\tilde{\epsilon}_X]_{e\mu}$ and the medium baseline to $[\tilde{\epsilon}_X]_{e\beta}$. This is the opposite what happens for the amplitude of oscillations. 

We can compute the equivalent of Jarskolg invariant for the BSM scenario as described in Appendix \ref{invariants}. From Eq.~(\ref{eq:CP1}) and Eq.~(\ref{eq:CP2}) 
we can read that 
\begin{align}
   J_{e\mu}^{\rm CP} =& \Im{[\tilde{\epsilon}_X]_{e\mu}}c^3_{13} s_{2\theta_{12}},
   \label{eq:CP_solar} \\
   J_{e\tau}^{\rm CP}=& \Im{[\tilde{\epsilon}_X]_{e\tau}}s_{2\theta_{13}}.
   \label{eq:CP_atm}
\end{align} 
As discussed before in the standard neutrino oscillation scenario, when the initial and final neutrino are equal there is no CP violation. Then, if there is any evidence  of $ J_{e\alpha}^{\rm CP}\neq 0$ $(\alpha=\mu,\tau)$ it will signalling new physics. 

\subsection{Flavour conversion in the sun}
\label{sec:sun}

To arrive at the expressions for the effect of NSI in the solar neutrino data,
we return to Eq.~(\eqref{eq:rates}) and~\eqref{eq:vklpx}. Unlike reactor neutrinos,
the solar neutrino flux arrives at the detectors as an incoherent admixture of
mass eigenstates, so the sum in Eq.~\eqref{eq:rates} becomes:
\begin{equation}
    \frac{R_{\alpha\beta}}{\phi_\alpha^{\rm SM}\sigma_\beta^{\rm SM}}=\sum_{kl}\delta_{kl}\left[\widetilde{V}_{\alpha}^{kl}(p_X)\right]
    \times \left[V^{kl}_{\beta}(d_X)\right]^*.
    \label{eq:ratessun}
\end{equation}
Besides, we have to take into account the mixing angle in matter at the production point. Writing explicitly:
\begin{eqnarray}
  \frac{R_{\alpha\beta}}{\phi_\alpha^{\rm SM}\sigma_\beta^{\rm SM}} &=& \sum_k
  (|\widetilde{U}_{\alpha k}|^2+2\Re{p_{\rm XL}(\epsilon_X \widetilde{U})_{\alpha k}^*\widetilde{U}_{\alpha k}}+|p_{\rm XX}(\epsilon_X \widetilde{U})_{\alpha k}|^2) \nonumber \\
  &&\times\left(|U_{\beta k}|^2+2\Re{d_{\rm XL}(\epsilon_X U)_{\beta k}^*U_{\beta k}}+|d_{\rm XX}(\epsilon_X U)_{\beta k}|^2\right)
\label{eq:ratessun2}
\end{eqnarray}
where $\widetilde{U}$ denotes the mixing matrix in the production point inside the Sun. It is clear from Eq.~\eqref{eq:ratessun2} that solar neutrinos has no sensitivity to CP violation effects induced by NSI.

As a consistency test, it is straightforward, although cumbersome, to check that Eq.~\eqref{eq:ratessun2} reduces to Eq.~\eqref{eq:rates_as_prob} taking the average over the oscillation terms in  Eq.~\eqref{eq:rates_as_prob} and making $\tilde{U}\rightarrow U$ in Eq.~\eqref{eq:ratessun2}. However, it is clearer for solar neutrinos to emphasize the modifications induced by NSI on the initial fluxes and cross-sections.

For instance, for the electron neutrino flux created at the Sun, it is possible to write how such flux is shared between the mass eigenstates:

\begin{equation}
\Phi(\nu_k) =\phi(\nu_e)\left(P_{ek}^{\rm SM}
+2\Re{p_{\rm XL}(\epsilon_X \widetilde{U})_{\alpha k}^*\widetilde{U}_{\alpha k}}+|p_{\rm XX}(\epsilon_X \widetilde{U})_{\alpha k}|^2) \right)
\end{equation}
and, similarly, the effective cross-section of detecting a $\beta$-flavour neutrino on Earth from a mass eigenstate $k$ can be written as:
\begin{equation}
\sigma_\beta=\sigma_\beta^{SM}\left(P_{k\beta}^{\rm SM}+2\Re{d_{\rm XL}(\epsilon_X U)_{\beta k}^*U_{\beta k}}+|d_{\rm XX}(\epsilon_X U)_{\beta k}|^2\right)
\end{equation}

With these expressions, we can implement the NSI effects as corrections on
production probabilities of mass eigenstates in the Sun and corresponding probabilities of detecting a mass eigenstate at the detectors. However, it is essential to note that these are not actual probabilities, not adding to $1$ when all states are considered.

Using the same ansatz given by Eqs.~\eqref{eq:nsinew1} to Eq.~\eqref{eq:mynsi} (but for neutrinos), we present in the following the corrections on such probabilities.

For solar production of mass eigenstates and $\tilde{\epsilon}_{e\mu}$, we have:
\begin{eqnarray}
P_{e1}&\rightarrow& P_{e1}^{\rm SM}-2\Re{[\widetilde\epsilon_X]_{e\mu}} p_{\rm XL} \widetilde{s}_{12}\widetilde{c}_{12}c_{13}+|[\widetilde\epsilon_X]_{e\mu}|^2p_{\rm XX}\widetilde{s}^2_{12} \nonumber \\
P_{e2}&\rightarrow& P_{e2}^{\rm SM}+2\Re{[\widetilde\epsilon_X]_{e\mu}} p_{\rm XL} \widetilde{s}_{12}\widetilde{c}_{12}c_{13}+|[\tilde\epsilon_X]_{e\mu}|^2p_{\rm XX}\widetilde{c}^2_{12} \nonumber \\
P_{e3}&=&P_{e3}^{\rm SM}
\end{eqnarray}
 
For $\tilde{\epsilon}_{e\tau}$, we have:
\begin{eqnarray}
P_{e1}&\rightarrow& P_{e1}^{\rm SM}-2\Re{[\tilde\epsilon_X]_{e\mu}} p_{\rm XL} \tilde{c}^2_{12}s_{13}c_{13}+|[\tilde\epsilon_X]_{e\mu}|^2p_{\rm XX}\tilde{c}^2_{12}s^2_{13} \nonumber \\
P_{e2}&\rightarrow& P_{e2}^{\rm SM}-2\Re{[\tilde\epsilon_X]_{e\mu}} p_{\rm XL} \tilde{s}^2_{12}s_{13}c_{13}+|[\tilde\epsilon_X]_{e\mu}|^2p_{\rm XX}\tilde{s}^2_{12}s^2_{13} \nonumber \\
P_{e3}&=&P_{e3}^{\rm SM}+2\Re{[\tilde\epsilon_X]_{e\mu}} p_{\rm XL}c_{13}+|[\tilde\epsilon_X]_{e\mu}|^2p_{\rm XX}\tilde{c}^2_{13}
\label{faca}
\end{eqnarray}
and for the detection on Earth we obtain similar expressions, with $d_{XL}\rightarrow p_{XL}$ and $\tilde{U}\rightarrow{U}$.

We can note that for $\tilde{\epsilon}_{e\tau}$ the corrections on the probabilities involving the two first families, which are the most relevant for solar neutrinos, are suppressed by the small value of $\theta_{13}$. Indeed, we explicitly checked that the constraints on $[\tilde\epsilon_X]_{e\tau}$ from solar neutrino data are much weaker than those coming from reactor neutrino experiments.

As stated above, we assume that the probabilities in solar neutrino flux can be treated classically since the flux is an incoherent sum of mass eigenstates. However, if the neutrinos arrive at the detector during the night, the coherence between mass eigenstates is reestablished, and flavor oscillation can be probed. We developed expressions to include NSI on neutrino regeneration on Earth and explicitly checked that its effect on the constraints is marginal.

\subsubsection{Solar neutrino production and detection factors}
\label{proddectsun}
For the solar neutrinos, we assume mainly Gamow-Teller production. As in the reactor case, we use the Lagrangian of Eqs. \eqref{eq:NSIL}, and we always consider each type of non-standard interaction per time.

\begin{table}[h!]
\centering
\begin{tabular}{|l|cc|l|cc|}
 & scalar & tensor & & scalar & tensor \\ \hline 
$p_{\rm XL}$  & $0$ & $+\dfrac{g_T}{g_A}\dfrac{m_e}{E_e}$  & $p_{\rm XX}$  & $\dfrac{g^2_S}{3g_A^2}$ & $\dfrac{g^2_T}{g_A^2}$
\\&&&&&\\$d_{\rm XL}$  &  $-\dfrac{g_Sg_V}{g_V^2+3g_A^2}\dfrac{m_e}{E_e}$ & $\dfrac{3g_Ag_T}{g_V^2+3g_A^2}\dfrac{m_e}{E_e}$ &
$d_{\rm XX}$  & $\dfrac{g_S^2}{g_V^2+3g_A^2}$ & $\dfrac{3g_T^2}{g_V^2+3g_A^2}$
\end{tabular}
\caption{The same as Table \ref{tab:factorsractors} but now for production and detection factors for solar neutrinos.
}
\label{tab:factors_sun}
\end{table}

In general, the neutrino production in the solar core comes from several chain reactions including proton-proton collisions and beta decays. In this work, we restrict ourselves to beta decays (in the solar case, the $\beta^+$). As a result, any other production mechanism will be the same as in the standard model. The detection, will be considered as inverse beta decays, $\beta^-$:

\begin{equation}
    \begin{aligned}
        \beta^+:\;&^A_ZN_i \rightarrow ^A_{Z-1}N_f+e^++\nu_e, &\text{(production)}\\
        \text{inv. }\beta^-:\;&^A_ZN_i+\nu_e\rightarrow e^-+  ^A_{Z+1}N_f,&\text{(detection)}
    \end{aligned}
\label{eq:reactions}
\end{equation}
The $p_{\rm XY}$ and $d_{\rm XL}$, Eq. \eqref{eq:std_nsi_ratio}, for the sun are presented in table \ref{tab:factors_sun}. We follow the same notation as in table \ref{tab:factorsractors}.

\section{Analysis details}
\label{sec:analysis}

\subsection{Medium baseline reactors}
\label{mbr}

Here we explore the reactors medium baseline experiments: Daya Bay~ \cite{PhysRevLett.121.241805}, RENO~\cite{PhysRevLett.121.201801} and Double Chooz~\cite{DoubleChooz:2019qbj}, those are considered Medium Baseline Reactors (MBR) and are sensitive to the $\Delta m^2_{31}$. For that case, we use the GlobesFit code~\cite{Berryman_2021} modified to include non-standard interactions. For further details on the experimental setup see Ref.~\cite{Berryman_2021}. For the MBR case, we adopt the following $\chi^2$ distribution:

\begin{equation}
    \chi^2_{\rm MBR}= \sum_{{\rm exp}=\{{\rm DB,DC,RENO}\}}(\chi^{\rm shape}_{\rm exp})^2+(\chi^{\rm rate}_{\rm exp})^2+\frac{(1-\alpha)^2}{\sigma_a^2},
    \label{eq:chi_MBR}
\end{equation}
where,
\begin{equation}
    (\chi^2)_{\rm DB}^{\rm shape}= \sum_{k=\{\rm EH2,EH3\}}\sum_{i,j}^{N_{\rm  DB}}(d_i^k-n_i^k)(V^{-1}_{\rm DB})_{ij}(d_j^k-n_j^k),
    \label{eq:chi_ratios}
\end{equation}
here, $d^k_i$ is the released data on the energy bin $i$ (over $N_{\rm  DB}=52$ energy bins) for the ratio between the number events in the experimental hall $k=1,2,3$ (EHk) with EH1, $n_i^k$ is the corresponding simulated value. Here $V_{\rm DB}$ is the covariance matrix for the Daya Bay shape analysis. And

\begin{equation}
    (\chi^2)_{\rm DB}^{\rm rate}= \sum_{i,j}^{N_{\rm DB}}(d_{0,i}^k-(1-\alpha)n_{0,i}^k)(W^{-1}_{\rm DB})_{\rm ij}(d_{0,j}^k-(1-\alpha)n_{0,j}^k),
    \label{eq:chi_total_rates}
\end{equation}
where the variable have similar meanings as in the shape analysis case, with the difference that $d_{0,i}(n_{0,i})$ ($i$ runs over $N_{\rm  DB}=8$  bins for different running periods) is the ratio between the total number of events at AD1, AD2, AD8 and AD3 divided by the standard neutrino oscillation prediction case. The parameters $\alpha$ controls the normalization error, $\sigma_a=0.025$, that we assume to be fuel independent.
For Double Chooz (DC) ({$N_{\rm  DC}=26$} data points in the shape analysis and 4 data points in the rate) and  RENO ($N_{\rm  RENO}=25$ data points in the shape analysis and 8 data points in the rate) we consider near and far detectors ratio. The $\chi^2$ is the same as in Eqs. \eqref{eq:chi_ratios} and \eqref{eq:chi_total_rates} but $d_i^k$ is the far/near ratios and $d_{0,i}^k$ are the ratios in the near detector with the standard model prediction.
\subsection{KamLand experiment}

In this section, we describe the setup we used to simulate KamLand experiment \cite{kamland2013}. The KamLand experiment consists of 1 kton of highly purified liquid scintillator detector in Japan that collects signals of inverse beta decay of neutrinos coming from different reactors in Japan. As the average distance from the reactors and the KamLand detector is of the order of 180 km, and the neutrino energies are around 3 MeV, the experiment is sensitive to the solar neutrino mass squared difference $\Delta m_{21}^2$. 

The KamLand detector measures the event rates for the anti-neutrino signal from 2002 to 2007. The total exposure of the experiment was $2.44 \times 10^{32}$ proton-yr. We consider the KamLand number of neutrino events in a bin as given by:
\begin{equation}
n_i \propto \epsilon_{i}\int_{E_i^{\rm min}}^{E_f^{\rm max}}dE_{e} \int_{E_{\rm th}}^{\infty}dE_\nu \left(\sum_j W_j  \left(\frac{R_{\overline{\nu}_e}\overline{\nu}_e}{\phi_e^{\rm SM}\sigma_e^{\rm SM}}\right)_{(E_\nu,L_j)}
\right)\frac{d\Phi(E_\nu)}{dE_\nu}\frac{d\sigma(E_\nu)}{dE_{e}} R(E_\nu,E_R),
\end{equation}
where the $\epsilon_i$ is the efficiency of the bin, $E_\nu$ is the neutrino energy, $E_{\rm th}$ is the threshold energy for the neutrino interaction, $E_{e}$ is the positron energy, $R(E_\nu,E_R)$ is the energy resolution function, $W_j$ is the power of the j-th reactor, $\left(\frac{R_{\overline{\nu}_e}\overline{\nu}_e}{\phi_e^{\rm SM}\sigma_e^{\rm SM}}\right)_{(E_\nu,L_j)}$ will be given by the oscillation rate Eq. \eqref{eq:rates}, $\frac{d\sigma(E_\nu)}{dE_{e}}$ is the detection differential cross-section and the anti-neutrino flux at the detector is given by 
\begin{equation}
\frac{d\Phi(E_\nu)}{dE_\nu} = \frac{\sum_j F_j\dfrac{d\phi^{(j)}_\nu(E_\nu)}{dE_\nu}}{4\pi\sum_iW_i L_i^2},
\end{equation}
where $F_j$ are the fraction of the reactor fuel given by KamLand collaboration $(0.567:0.078:0.298:0.057)$ for 
($^{235}$U : $^{238}$U : $^{239}$Pu : $^{241}$Pu). The neutrino flux comes from \cite{Estienne_2019} and the baselines $L_i$ and power $W_i$ for each reactor comes from \cite{SUEKANE2006106}. The efficiencies are given by $\epsilon_i$ and extracted from \cite{kamland2013}. The anti-neutrino cross-section comes from \cite{Vogel_1999} and the neutrino reconstruction function was used \cite{HUBER2011360} using a 6.4\%$\sqrt{E_\nu (\rm MeV)}$ error given by \cite{kamland2013}.

For the statistical analyse of KamLand (KL), we calculate the following $\chi^2$:

\begin{equation}
    \chi^2_{\rm KL} = \sum_{i}\frac{\Big(d_i-n_i(a_1,a_2)-b_i(a_3,a_4)\Big)^2}{d_i}+\sum_i\frac{a_i^2}{\sigma_i^2}
    \label{eq:chi_KL}
\end{equation}
where $d_i$ is the data extracted by summing the number of events of the three  KamLand phases of Ref. \cite{kamland2013}, $b_i$ is the background also extracted from Ref. \cite{kamland2013}, $\sigma_i$ are systematic errors, we used $\sigma_1=0.05$(signal normalization error), $\sigma_2=0.02$ (signal energy error), $\sigma_3=0.08$ (background normalization error) and $\sigma_4=0.02$ (background energy error). Also, the number of events given the calibration and normalization errors will be the same as in the Ref.~\cite{Huber:2007globes}:
\begin{equation}
    n_i(a,b) = (1+b)(1+a)\left[\big(n_{i+1}-n_i\big)\big(\delta(b)-i\big)+n_i\right],
\end{equation}
here, $\delta(b) = b(i+t_0+0.5)+i$ where $t_0=N_{\rm bins}E_{\rm min}/(E_{\rm max}-E_{\rm min})$. For each $\chi^2_{\rm KL}$ we minimize over the nuisance parameters $a_i$.

\subsection{Solar experiments}

For the solar neutrino data, we perform a statistical analysis following the same procedure presented in~\cite{deHolanda:2008nn,deHolanda:2003nj}. The solar neutrino data used are:

\begin{itemize}
    \item the full spectral data from Super-Kamiokande phases I, III and IV~\cite{Hosaka:2005um,Fukuda:2002pe,Abe:2016nxk};
    
    \item the combined analysis of all three SNO phases~\cite{Aharmim:2011vm};
    
    \item Borexino results~\cite{Agostini:2018uly};
    
    \item combined Gallex+GNO~\cite{Kaether:2010ag}  and SAGE~\cite{Abdurashitov:2009tn} data;
    
    \item Homestake results~\cite{Cleveland:1998nv}. 
\end{itemize}

\subsection{Global picture}
\label{sec:global}
For our analysis of the global electron disappearance we can adopt the uncorrelated test statistics summing the $\chi^2$ functions, Eqs. \eqref{eq:chi_MBR},\eqref{eq:chi_KL} and $\chi^2_{\rm Sun}$, for MBR, KamLand and solar data:

\begin{equation}
    \chi^2_{\rm global}=\chi^2_{\rm MBR}+\chi^2_{\rm KL}+\chi^2_{\rm Sun}.
    \label{eq:global_chi}
\end{equation}

We scan the $\chi^2$ function over all the parameters that can be sensitive in each of the experiments listed here. The scan was stored separated in large tables for MBR, KamLand, and Solar experiments,  later combined in a larger table. In the standard neutrino oscillation, the MBR experiments are sensitive to $\Delta m^2_{31}$ and $\theta_{13}$. When non-standard interactions are present, in addition, we have the Re$[\tilde{\epsilon}_X]_{e\alpha}$ and Im$[\tilde{\epsilon}_X]_{e\alpha}$ resulting in 4 parameters. As was discussed in Sec. \ref{proddectmbr}, the effect of inclusion of the Im$[\tilde{\epsilon}_X]_{e\mu}$ makes the MBR also sensitive to the $\Delta m_{21}^2$ and $\tan\theta_{12}$ through Eq. \eqref{eq:CP1}, elevating the number of parameters to 6. In that case, we fixed $\theta_{12} = 33.4^o$ (in order to reduce the computational time) and vary only $\Delta m_{21}^2$ reducing the number of parameters to 5.

The KamLand and solar experiments are sensitive to the $\theta_{12}$, $\Delta m^2_{21}$ and $\theta_{13}$ parameters in standard neutrino oscillation. With the presence of the new interactions, the real and imaginary parts of the couplings, Re$[\tilde{\epsilon}_X]_{e\alpha}$ and Im$[\tilde{\epsilon}_X]_{e\alpha}$, will be sensitive in those experiments rising the number of parameters to 5. Furthermore, as was discussed in Sections \ref{proddectmbr} and \ref{proddectsun} the $[\tilde{\epsilon}_X]_{e\tau}$ is suppressed by a small value of $\theta_{13}$, measurement that comes from MBR experiments. So, for solar experiments, only the $[\tilde{\epsilon}_X]_{e\mu}$ parameter will be considered. For KamLand experiments, we still consider the $[\tilde{\epsilon}_X]_{e\tau}$ parameter in the final analysis but we check that it bring minimal changes in the final result.

In brief, in the global analysis we considered all the parameters: $\Re [\tilde{\epsilon}_X]_{e\alpha}$, $\Im [\tilde{\epsilon}_X]_{e\alpha}$, $\theta_{12}$, $\theta_{13}$, $\Delta m^2_{21}$ and $\Delta m^2_{31}$, a total of 6 parameters. The global analysis was made by summing the $\chi^2$ functions and interpolating the values of the KamLand, Solar, and MBR tables. As we will see, all the parameters play an important role in the analysis.

\section{Results}
\label{sec:results}
In this section, we describe the result of our analysis of tensor and scalar interactions in the neutrino and anti-neutrino disappearance in the experiments considered in Sec. \ref{sec:analysis}. It was explored a total of four independent analysis in this work, for non-zero $[\tilde{\epsilon}_T]_{e\tau}$, $[\tilde{\epsilon}_S]_{e\tau}$, $[\tilde{\epsilon}_T]_{e\mu}$ and $[\tilde{\epsilon}_S]_{e\mu}$.  We use the $\chi^2$ statistics described in Eq. \eqref{eq:global_chi} whose 
the minimum value for each analysis case is shown in Tab.
(\ref{tab:summary}) together with the number of data points in each experiment. We have found that all the experiments presents a $\chi^2/n_{\text{data}}<1$ in all scenarios individually or combined for any of analysis cases studied here. We use a total of 280 data points, 149 from solar experiments, 17 from KamLand, and 114 for the MBR (shape+rate analysis).

\begin{table}[hbt]
\centering
\begin{tabular}{|c|l|l|l|l|l|c|}
\hline
 & S.O. & $[\tilde{\epsilon}_S]_{e\tau}$ & $[\tilde{\epsilon}_T]_{e\tau}$ & $[\tilde{\epsilon}_S]_{e\mu}$ & $[\tilde{\epsilon}_T]_{e\mu}$ & $n_{\rm data}$\\ \hline
$\left(\chi^2_{\rm Sun}\right)_{\rm min}$ & 134.0 & 134.0 & 134.0 & 133.9 & 132.0 & 149\\ \hline
$\left(\chi^2_{\rm KAMLAND} \right)_{\rm min}$ & 15.0  & 14.9 & 14.5 & 15.0 & 14.6 & 17 \\ \hline
$\left(\chi^2_{\rm MBR}\right)_{\rm min}$& 87.0 &  84.1 & 85.1 & 83.9 & 82.5 & 114 \\ \hline
$\left(\chi^2_{\rm global}\right)_{\rm min}$& 242.7 & 240.3 & 241.3 & 242.3 & 239.9 & 280\\ \hline
$\Delta\overline{\chi}^2_{\rm PG}$ (p-value)& 15.7\% & 29.0\% & 26.0\% & 38.3\% & 29.0\% & |\\ \hline
\end{tabular}
\caption{Summary of the analysis, here $\left(\chi^2_{X}\right)_{\rm min}$ is the local minimum $\chi^2$ for experiment X=medium baseline reactor experiments (MBR), KamLand and solar experiments for standard neutrino oscillations (S.O.) and for different NSI scenarios.
Here, $\Delta\overline{\chi}^2_{\rm PG}$ as defined in Eq. \eqref{eq:PG}. The number of data points, $n_{\rm data}$,  are shown in the last column.}
\label{tab:summary}
\end{table}
Moreover, in order to check the parameter consistency over different data sets, we also compute the parameter goodness (PG) of fit \cite{Maltoni_2003} for each case in the analysis, by using the:
\begin{equation}
   \Delta\overline{\chi}^2_{\rm PG}=  (\chi^2_{\rm global})_{\rm min} -({\chi^2_{\rm MBR})_{\rm min}-(\chi^2_{\rm KL})_{\rm min}-(\chi^2_{\rm Sun}})_{\rm min},
    \label{eq:PG}
\end{equation}
where the {$\Delta\overline{\chi}^2_{\rm PG}$} is distributed as a $\chi^2$ statistics with $P_c=\sum_r P_r-P$ d.o.f., where $P_r$ is the number of parameters for the experiment $r$ and $P$ is the total number of parameters in the global analysis. For example, for the $[\tilde{\epsilon}_X]_{e\mu}$ we have $P_c=5+5+5-6=9$ d.o.f. (5 for KamLand, 5 for the sun and 5 for MBR) and for $[\tilde{\epsilon}_X]_{e\tau}$, we have $P_c=\sum_r P_r-P=5+3+4-6=6$ d.o.f. (4 for KamLand, 3 for the sun and 5 for MBR). From the results of the parameter goodness o the fit, PG, listed in the last line of Tab.( \ref{tab:summary}), we found a good agreement in the PG for all the scenarios, with an small improvement of the agreement between data and model when NSI is present.

\begin{figure}[hbt]
    \centering
    \includegraphics[scale=0.8]{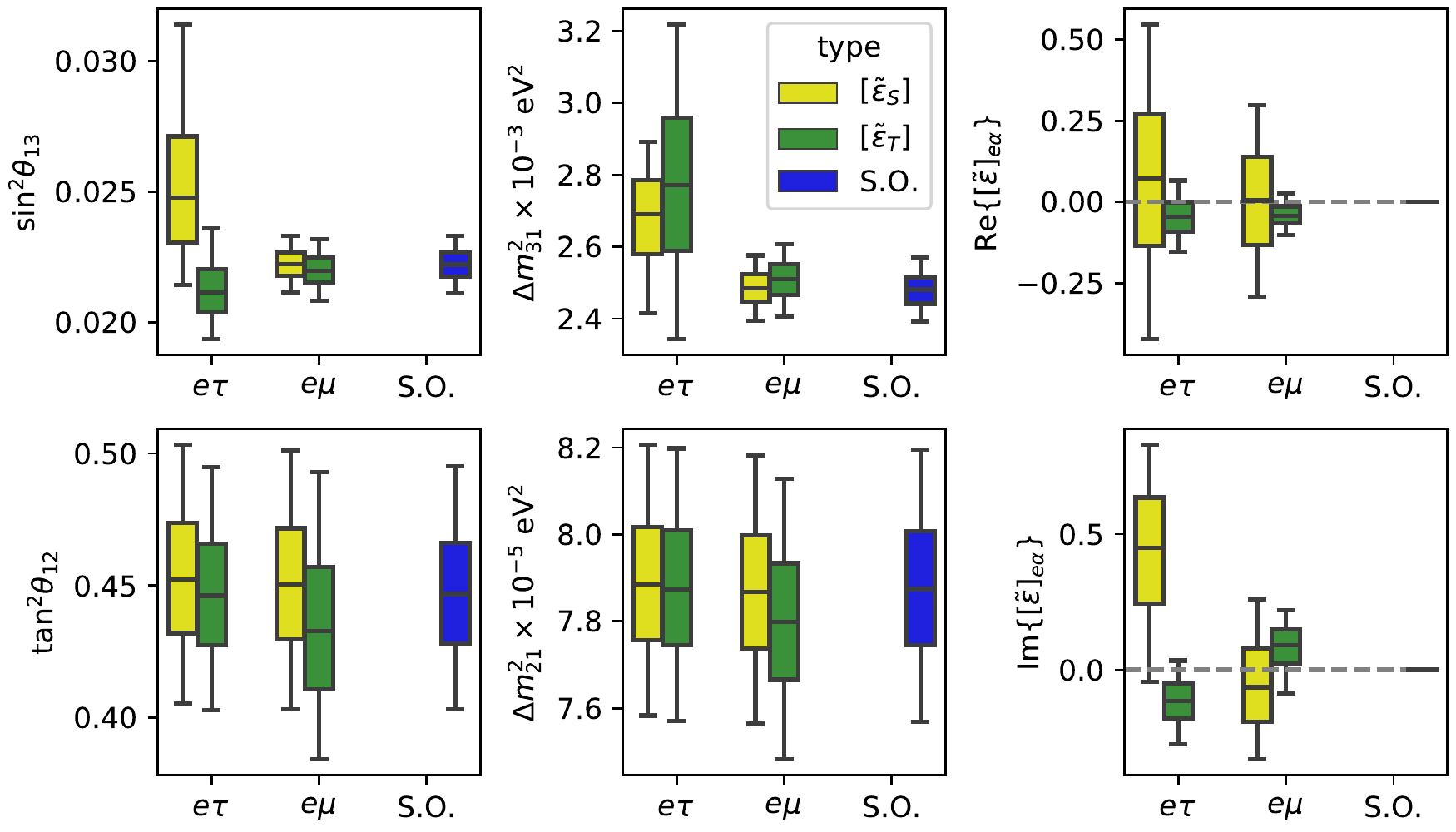}
    \caption{The box-plot for the parameters distributions. We show the median, the first and third quartile boxes and the 5\% and 95\% percentiles for the whiskers ($\approx$90\% C.L.).}
    \label{fig:global_lom}
\end{figure}

As expected, the standard neutrino oscillation (S.O.) produces a fit that is in conformity with the data. The global minimum is  $\chi^2$: $\left(\chi^2_{\rm global}\right)_{\rm min}=242.6$ as shown in Table~(\ref{tab:summary}). In the standard neutrino oscillation, 4 parameters were considered under the normal ordering hypothesis: $\theta_{13}$, $\theta_{12}$, $\Delta m^2_{31}$ and $\Delta m^2_{21}$. Individually, each of those experiments is in agreement with the data ($\left(\chi^2_{\rm KL}\right)_{\rm min}=15.0$, $\left(\chi^2_{\rm Sun}\right)_{\rm min}=134.0$ and $\left(\chi^2_{\rm KL+Sun}\right)_{\rm min}=155.1$ ), the solar and KamLand are sensitive to $\theta_{12}$, $\Delta m^2_{21}$ and can alone put a weak maximum size for $\theta_{13}$. The MBR experiments are either in agreement with data, leading to $\left(\chi^2_{\rm MBR}\right)_{\rm min}=87.0$ and are sensitive to $\theta_{13}$ and $\Delta m^2_{31}$. In Fig. \ref{fig:global_lom} we have shown the boxplot using the whiskers at a range of 90 \% C.L. for the mixing parameters and the squared mass differences in the blue box. Our range is in agreement with the global analysis of neutrino experiments like NUFIT~\cite{Esteban:2020cvm}.

As previously discussed in Sec. \ref{proddectmbr}, \ref{proddectsun} and \ref{sec:global}, when considering NSI we noticed that KamLand and Solar experiments are more sensitive to $[\tilde{\epsilon}_X]_{e\mu}$ than in MBR. On the other hand, the MBR sensitivity to $[\tilde{\epsilon}_X]_{e\tau}$ is much more robust than in the KamLand and solar experiments, turning their contribution negligible to the global analysis. Those sensitivity associations can be graphically seen in 
Fig.~(\ref{fig:graph_param}). Now, let us give a summary of comparing each NSI case with the standard neutrino oscillation scenario.
For each case we have found that the global minimum in the case of NSI,$(\chi^2_{\rm global})_{\rm min}$  is lower than the standard neutrino oscillation global minimum, $((\chi^2_{\rm global})_{\rm S.O.})_{\rm min}$,

\begin{itemize}
    \item Scalar $[\tilde{\epsilon}_S]_{e\tau}$: for this parameter we found a $(\chi^2_{\rm Global})_{\rm min}-((\chi^2_{\rm Global})_{
    \rm SO})_{\rm min} = 2.99$ that corresponds to a $1.73\sigma$ preference for non-zero NSI. The parameters goodness of fit has the p-values of 29\% showing that the best-fit is in agreement with the data over the different data sets:
    \item Scalar $[\tilde{\epsilon}_S]_{e\mu}$: we found  $[\tilde{\epsilon}_T]_{e\tau}$: we found $(\chi^2_{\rm global})_{\rm min}-((\chi^2_{\rm global})_{
    \rm SO})_{\rm min}= 2.40$. The same conclusion as before  with  a $1.55\sigma$ preference for non-zero NSI and the   p-values of 38\%;
    \item Tensor $[\tilde{\epsilon}_T]_{e\tau}$: we found $(\chi^2_{\rm global})_{\rm min}-((\chi^2_{\rm global})_{
    \rm SO})_{\rm min} = 1.99$. The same conclusion as before  with  a $1.41\sigma$ preference for non-zero NSI and the   p-values of 26\%;
    \item Tensor $[\tilde{\epsilon}_T]_{e\mu}$: we found  $[\tilde{\epsilon}_T]_{e\tau}$: we found $(\chi^2_{\rm global})_{\rm min}-((\chi^2_{\rm global})_{
    \rm SO})_{\rm min} = 2.89$. 
    The same conclusion as before  with  a $1.70\sigma$ preference for non-zero NSI and the p-values of 29\%.
\end{itemize}
Finally, in Tab. \ref{tab:summary} and in Fig \ref{fig:global_lom}, we summarize the relevant statistical results of this work. The $[\tilde{\epsilon}_X]_{e\tau}$ and $[\tilde{\epsilon}_X]_{e\mu}$ results will be presented in details in the next sections.

\subsection{The $[\tilde{\epsilon}_X]_{e\tau}$ analysis}
\label{etauanalysis}

\begin{figure}[hbt]
    \centering
    \includegraphics[scale=0.84]{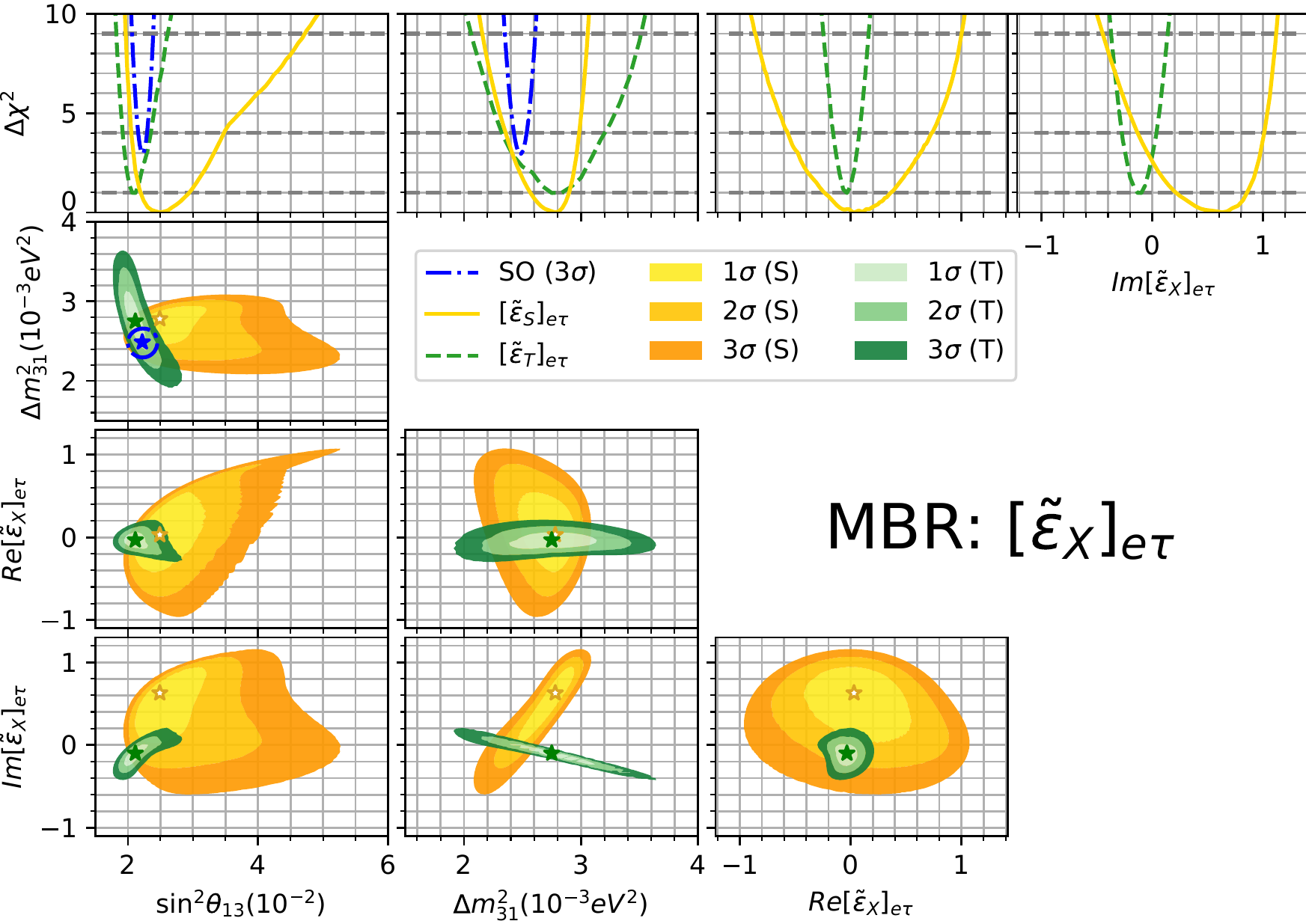}
    \caption{Upper panel: The 1D projection of the $\Delta \chi^2\equiv \chi^2-\chi^2_{\rm min}$ of medium baseline reactors (MBR) in the presence of $[\tilde{\epsilon}_X]_{e\tau}$ for the different parameters. Lower panels  the contour levels for oscillations parameters and the BSM couplings at $1\sigma$, $2\sigma$ and $3\sigma$. In green (yellow) is the NSI tensor (scalar) case and in dash-dotted black is standard neutrino oscillation case. }
    \label{fig:MBL-et}
\end{figure}

For the $[\tilde{\epsilon}_X]_{e\tau}$, the leading contributions to constrain this parameter to come from MBR experiments. In KamLand and solar experiments, the $[\tilde{\epsilon}_X]_{e\tau}$ NSI parameters are always suppressed by a small $\theta_{13}$, as we can see in eq.~(\ref{eq7} for KamLand experiment and in eq.~(\ref{faca} for solar experiments. The suppression of $[\tilde{\epsilon}_X]_{e\tau}$ contribution lead to little changes in the global analysis. Hence, we can consider only MBR experiments for the analysis of $[\tilde{\epsilon}_X]_{e\tau}$ parameters. This is displayed graphically in fig.~(\ref{fig:graph_param}).
The resulting medium baseline reactors $\chi^2$ functions, Eq. \eqref{eq:chi_MBR}, are shown in Fig. \ref{fig:MBL-et} for scalar and tensor interactions.

In the upper panel of Fig. \ref{fig:MBL-et} we show the difference between the $\chi^2$ and local minimum for MBR analysis, $\Delta \chi^2\equiv \chi^2-\chi^2_{\rm min}$ for the scalar (S)  and tensor (T) interactions, where X=S,T and for the standard neutrino oscillation scenario respectively the yellow, green and dotted blue curve.
The value of $(\chi^2_{\rm min}$ for each case, scalar, tensor or standard neutrino oscillation  can be read from Table~(\ref{tab:summary}). The two leftmost upper panels, show respectively as a function of the $\sin^2 \theta_{13}$ and $\Delta m_{31}^2$. The difference between the heights of the three curves, scalar, tensor, and standard neutrino oscillation, is due to each one having a different minimum. There is a slight preference for scalar interaction compared with the others solutions.  

In the lowest panels, we show the 2D contours for each combination of parameters critical to the MBR analysis, and the contours reveal all the correlations between variables. For each pair of variables, we minimize the complementary parameter space. The yellow, green, and blue curves with different C.L. are shown respect for the scalar, tensor, and standard neutrino oscillation. 
The more notable feature is that the allowed region of the mixing angles and squared mass differences for the scalar and tensor interactions scenario is enlarged compared with standard neutrino oscillations. 

Subsequently, we will discuss in details each interaction:

\begin{itemize}
\item For scalar $[\tilde{\epsilon}_S]_{e\tau}$ interactions, the standard neutrino oscillation is disfavored by $1.7\sigma$ in comparison with the NSI scenario. In the $\Delta m^2_{31}\times \Re[\tilde{\epsilon}_S]_{e\tau}$ panel of Fig. \ref{fig:MBL-et}, we can see that a positive correlation between $\Delta m^2_{31}$ and the CP violation term (imaginary part of $[\tilde{\epsilon}_S]_{e\tau}$) drives the improvement in the fit leading to a large $\Delta m^2_{31} = 2.78\times10^{-3}$~eV$^2$ and a non-zero $\Im [\tilde{\epsilon}_S]_{e\tau} = +0.62$. The analysis also leads to a $3\sigma$ allowed region of $\sin^2\theta_{13}\approx 0.05$ due to a correlation of the $\sin^2\theta_{13}$ with the $\Re [\tilde{\epsilon}_S]_{e\tau}$ as can be seen in the $\sin^2 \theta_{13}\times \Re[\tilde{\epsilon}_S]_{e\tau}$ panel. 
 The range of the parameters can be obtained by $\Delta \chi^2=1$ in the Fig.~(\ref{fig:MBL-et}).
The values of the real and imaginary parts found for scalar interactions $[\tilde{\epsilon}_S]_{e\tau}$ are
(shown graphically in right panel of Fig.~(\ref{fig:global_lom}) by the yellow box),

\begin{equation}
    \begin{aligned}
        \Re [\tilde{\epsilon}_S]_{e\tau} = +0.03_{ -0.21}^{+0.40},  & & \Im [\tilde{\epsilon}_S]_{e\tau} = +0.62^{+0.23}_{-0.41}. 
    \end{aligned}
    \label{eq:lim_Set}
\end{equation}

\item In the tensor interaction case, the standard neutrino oscillation is disfavored by a $1.41\sigma$ statistical significance, see Fig. \ref{fig:MBL-et}. The preference, as in the scalar case was guided by the $\Im [\tilde{\epsilon}_T]_{e\tau}$ that is is negatively correlated with $\Delta m^2_{31}$ also leading to a larger value of $\Delta m^2_{31}=2.75\times 10^{-3}$ eV$^2$. Now the imaginary part is negative $ \Im [\tilde{\epsilon}_T]_{e\tau} = +0.12$, this comes from the fact that the interference term in the production see Table.~(\ref{tab:factorsractors}), is negative for tensor interactions and positive for scalar interactions. At $3\sigma$ in the tensor case the $\Delta m^2_{31}$ can reach values of $3.6\times 10^{-3}$ eV$^2$ and $\sin^2\theta_{13}=0.027$ upper limit at $3\sigma$. In the tensor case, the $\theta_{13}$ is also correlated with $\Im [\tilde{\epsilon}_T]_{e\tau}$ as can be seen in the $\theta_{13}\times \Im [\tilde{\epsilon}_T]_{e\tau}$ panel of Fig.~(\ref{fig:MBL-et}).
The values of the real and imaginary parts of the NSI parameter are  (shown graphically in right panel of Fig.~(\ref{fig:global_lom}) by the green box):

\begin{equation}
    \begin{aligned}
        \Re [\tilde{\epsilon}_T]_{e\tau} = -0.03_{ -0.06}^{+0.06},  & & \Im [\tilde{\epsilon}_T]_{e\tau} = -0.12^{+0.08}_{-0.10}. 
    \end{aligned}
    \label{eq:lim_Tet}
\end{equation}
\end{itemize}

For the mixing parameters and the squared mass difference, we shown in left and central panel of Fig.~(\ref{fig:global_lom}) the range of $\sin^2 \theta_{13}, \tan^2 \theta_{12}, \Delta m^2_{31}$ and $\Delta m^2_{21}$. Visually we can notice that $\sin^2 \theta_{13}$ and $\Delta m^2_{31}$ changed compared with the usual neutrino oscillation scenario. 

\subsection{The $[\tilde{\epsilon}_X]_{e\mu}$ analysis}
\label{emuanalysis}

\begin{figure}
\centering
\begin{subfigure}{\textwidth}
      \includegraphics[scale=0.85]{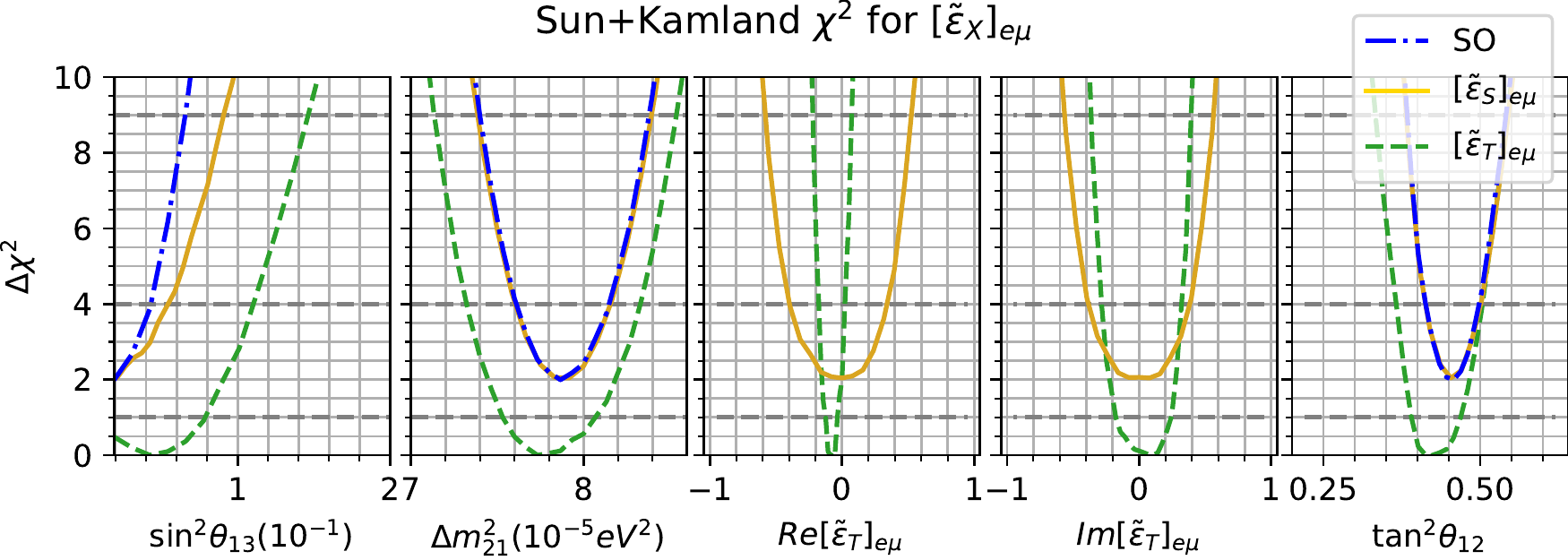}
    \caption{Here we show the $\Delta\chi^2$ as a functions for the $\sin^2\theta_{13}$, $\tan^2\theta_{12}$,$\Delta m^2_{21}$, $\Re[\tilde{\epsilon}_X]_{e\mu}$ and $\Im[\tilde{\epsilon}_X]_{e\mu}$ considering KamLand + solar data. The yellow (green)  curve we show the tensor (scalar) NSI scenario, and the dashed-dotted blue curve is the standard neutrino oscillation scenario (S.O.).}
  \label{fig:kamland+sol_chi}
\end{subfigure}\\
\begin{subfigure}{\textwidth}
\includegraphics[scale = 0.85]{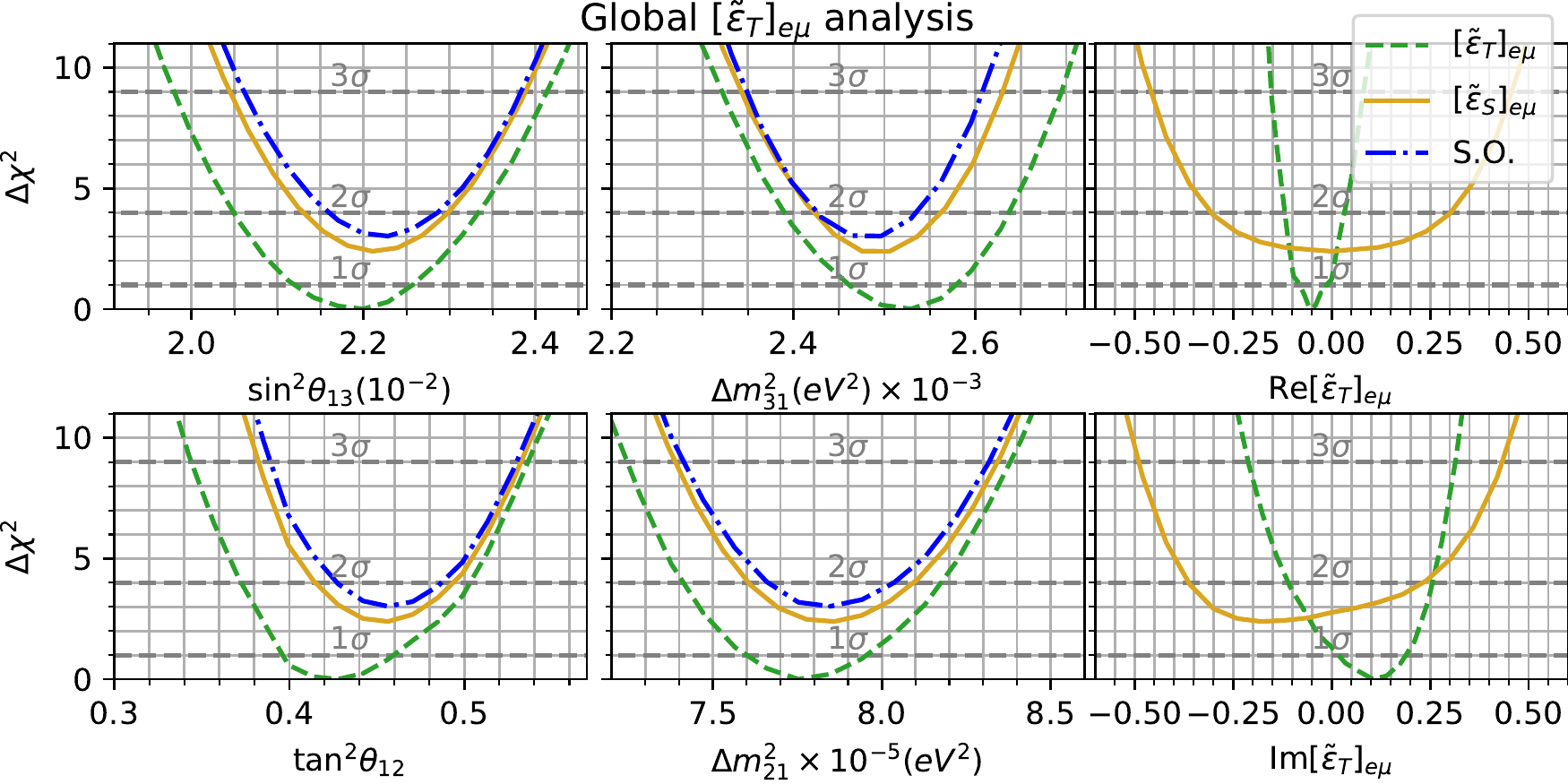}
 \caption{The same as before, but for the global analysis, MBR, KamLand and sun.}
  \label{fig:global}
\end{subfigure}
\caption{Results for $\Delta \chi^2$ for sun+KamLand and for the global analysis.}
\label{fig:test}
\end{figure}

The $[\tilde{\epsilon}_X]_{e\mu}$ parameter can be sensitive in MBR, KamLand and solar experiments. 
In Fig.~\ref{fig:kamland+sol_chi} we show the $
\Delta\chi^2$ function as mentioned before in Section~(\ref{etauanalysis}).  Comparing the  Fig.~(\ref{fig:kamland+sol_chi}) and ~(\ref{fig:MBL-et}) we have now that the tensor interaction is the best fit for neutrino data.

The inclusion of scalar interactions, the yellow curve of Fig. \ref{fig:kamland+sol_chi}, has negligible effects on the measurement of the standard neutrino oscillation parameters except for a slight change in the $\theta_{13}$. The preference of the data in the scalar interaction case is for zero NSI. In the tensor interaction case (green lines), there is a slight deviation of the $\tan^2 \theta_{12}$ parameter and $\Delta m^2_{21}$ with a $1.3\sigma$ preference for non-zero NSI. The solar experiments are where the $[\tilde{\epsilon}_T]_{e\mu}$ obtain its strongest bounds. More details of the solar analysis can see in appendix~\ref{sec:solar_analysis}.

In MBR, the effect appears not at the atmospheric scale terms as in the $[\tilde{\epsilon}_X]_{e\tau}$ parameter, but as a CP violation effect from the solar scale, see the discussion in section \ref{proddectmbr}. For the MBR analysis, the preference for non-zero NSI reaches $1.8\sigma$ for scalar interactions and $2.1\sigma$ for tensor interactions. This significance is mainly guided by the CP violation effect from the solar scale that appears as a completely new effect of the MBR experiments. More discussions on MBR analysis are presented in appendix \ref{sec:MBR_eu_analysis}.

The combined analysis (sun+KamLand+MBR) of scalar and tensor interactions are presented in Fig. \ref{fig:global}, where we use the same colors code as in Fig. \ref{fig:kamland+sol_chi}. We divide the discussion of the global analysis into a tensor and scalar interactions:

\begin{itemize}
\item In the presence of scalar $[\tilde{\epsilon}_S]_{e\mu}$ interactions, there are no effective changes in the statistical significance compared to the Standard Model. Also, the standard model parameters $\Delta m^2_{21}$ and $\theta_{12}$ have only small changes except for the $\theta_{13}$ as can be seen in Fig.(\ref{fig:global}). The main source of limits on the $[\tilde{\epsilon}_S]_{e\mu}$ comes from solar experiments.

From the global analysis of the $[\tilde{\epsilon}_S]_{e\mu}$ interactions, the limits on the NSI parameters are:
\begin{equation}
    \begin{aligned}
        \Re [\tilde{\epsilon}_S]_{e\mu} = 0.00\pm 0.25,  & & 
        \Im [\tilde{\epsilon}_S]_{e\mu} = -0.16_{-0.09}^{+0.31}. 
    \end{aligned}
    \label{eq:lim_Seu}
\end{equation}
which are the first limits on those parameters up to the publication of this work.

\item The case of tensor $[\tilde{\epsilon}_T]_{e\mu}$ interactions, remember that for the KamLand+Solar analysis, there is an improvement of $1.3\sigma$ in the statistics compared with the Standard Model (more information for the KamLand +Solar analysis are presented in the Appendix \ref{sec:solar_analysis}). In the MBR analysis, the significance is of $2.1\sigma$ guided by a CP violation term. In the global analysis case, there is an improvement of around $1.7\sigma$ compared to the standard neutrino oscillation scenario. Hence, the tensor $[\tilde{\epsilon}_T]_{e\mu}$ interaction can improve the fit of each experiment individually and when they are combined, which is very surprising. The global analysis results are presented in the green lines of Fig. \ref{fig:global}. The values of the NSI parameters for the $[\tilde{\epsilon}_T]_{e\mu}$ are:
\begin{equation}
    \begin{aligned}
        \Re [\tilde{\epsilon}_T]_{e\mu} = -0.05_{ -0.03}^{+0.04},  & & \Im [\tilde{\epsilon}_T]_{e\mu} = -0.13_{-0.07}^{+0.09}. 
    \end{aligned}
    \label{eq:lim_Teu}
\end{equation}
Those are the first limits on those parameters that point to a non-zero tensor interaction at $1.7\sigma$.
\end{itemize}

\begin{figure}[ht!]
    \centering
    \includegraphics[scale=0.85]{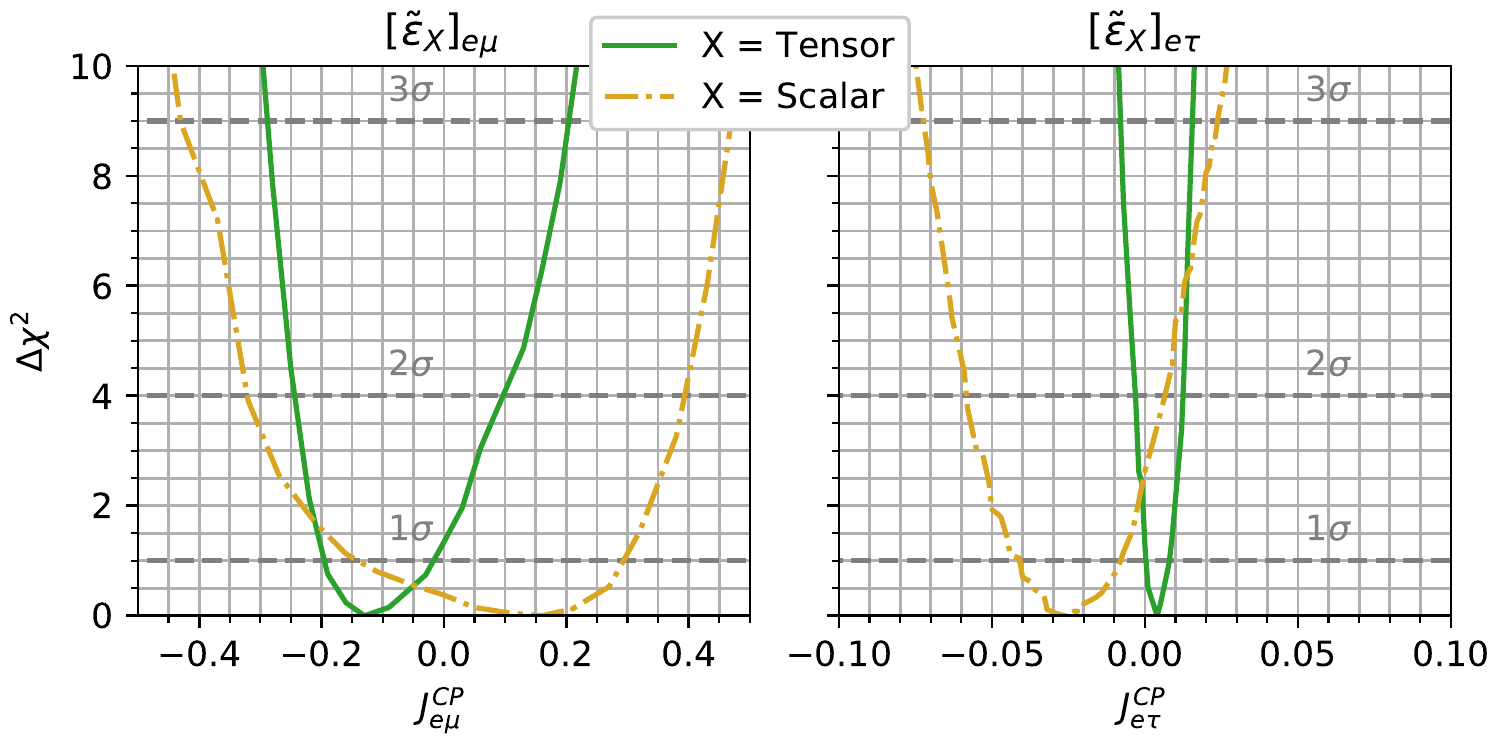}
    \caption{In this figure we show the $\Delta \chi^2$ as function of the CP-violation parameters, in the left (right) panel we show the $J_{e\tau}$($J_{e\mu}$) parameters of Eqs. \eqref{eq:CP_solar} and \eqref{eq:CP_atm}.}
    \label{fig:CPchi}
\end{figure}
\subsection{CP violation}

We have shown before the dependence on the imaginary part of NSI coupling, $[\tilde{\epsilon}_X]_{e\alpha}$, $\alpha=e,\mu$ and for both couplings X=S,T that made CP violation even in the case where the initial and final neutrino is the same. Nevertheless, the NSI couplings $[\tilde{\epsilon}_X]_{e\alpha}$ as well the mixing matrix $U_{\alpha i}$ it is not invariant by global re-phasing of fields. As discussed in the Appendix \ref{invariants}, the full rate Eq.~(\ref{eq:ratessun2}) it is invariant.  We compute the invariant combination in Eq.~(\ref{eq:CP_atm}), the quantities called $J_{e\mu}$ and $J_{e\tau}$ in the allowed region of mixing angle parameters and NSI couplings.

The CP violation phase appears in reactors and solar experiments. However, the CP violation effect appears only in reactors, as presented in Eqs. \eqref{eq:CP_solar} and \eqref{eq:CP_atm}. In this section, we quantify the effects of those CP violation terms. The resultant $\Delta \chi^2$ functions are presented in Fig. \ref{fig:CPchi}; in the left panel, we show the $\Delta \chi^2$ curve for the $e\mu$ interactions and in the right panel for the $e\tau$ interactions. The green curve are shown the tensor interactions in which the strength of CP is more constrained than the dashed-dotted yellow. There are two scenarios, $[\tilde{\epsilon}_T]_{e\mu}$ and $[\tilde{\epsilon}_S]_{e\tau}$, where a preference for non-zero CP violation in more than $1\sigma$ appears. For the tensor, $[\tilde{\epsilon}_T]_{e\tau}$ interactions, the constraint is the most restrictive. On the other hand, for the $[\tilde{\epsilon}_S]_{e\mu}$, the bounds are more relaxed.

We check that the lowest data point energy spectrum of Daya Bay was responsible for the improvement in the analysis and that such effect is explained as a CP violation effect. The significance of the effect is low,  only $1\sigma$, more experimental data are needed to support a non-statistical fluctuation interpretation. One experiment that could test such effect which much more statistics, can be the future designed 
JUNO experiment~\cite{JUNO_2016}.

\section{Conclusions}
\label{sec:conclusions}

We studied the effects on non-standard neutrino interaction in the production and detection of neutrinos and compared with the medium baseline reactors experiments,  Daya Bay, RENO, Double Chooz, KamLand, and solar experiments.

Usually, the non-standard neutrino interaction in the production and detection of neutrinos are assumed constants independent of energy. We follow the approach of ~\cite{Falkowski:2019kfn,Falkowski:2019xoe} to compute the changes in neutrino production and detection assuming scalar and tensor interactions.  This approach has interesting characteristics as the amplitudes of the oscillation are {\it now energy dependent}, due to the interference of scalar/tensor amplitude with standard model amplitudes. We decide to test this scenario for experiments with (anti)-electron neutrino in initial and in the final state as shown in Eq.(\ref{eq:rates}).  This scenario test the
flavor structure of electron disappearance experiments due the NSI   parameters $[\tilde{\epsilon}_X]_{e\mu}$ and $[\tilde{\epsilon}_X]_{e\tau}$. 
A side effect of this approach is that we will have {\it CP violation effects}, due to the imaginary part of non-standard couplings,  even when the initial and final neutrinos are the same. 

The summary of our results are shown in table (\ref{tab:summary}) and in the Figure (\ref{fig:global_lom}). The highlights of our scenario are a slight preference ($\sim 1 \sigma$) for non-zero CP violation effects as the best fit in all scenarios that we studied. 
For only reactor neutrino experiments, for the parameter $[\tilde{\epsilon}_X]_{e\tau}$ the results are in figure (\ref{fig:MBL-et}) that shown the increase in the allowed region of mixing angles and mass differences, mostly for $\sin^2 \theta_{13}$ and $\Delta m_{31}^2$. The inclusion of solar and KamLand experiments did not change much the results of reactor experiments.  
For the coupling $[\tilde{\epsilon}_X]_{e\mu}$, we have similar results, but here the KamLand and the solar neutrino experiments give good bounds on $[\tilde{\epsilon}_X]_{e\mu}$.

We test the parameter goodness-of-fit of all the parameters, see Eq.~\eqref{eq:PG}. We find an small improvement compared to the standard neutrino oscillation scenario as can be seen in Table \ref{tab:summary}, the p-value for the standard neutrino oscillation is $16\%$, for the non-standard scenario runs from 26\%  for $[\tilde{\epsilon}_T]_{e\tau}$ to $38\%$ in the $[\tilde{\epsilon}_S]_{e\mu}$ scenario. So, in all cases, we have an agreement for the parameters in different experiments.

We found new limits on NSI, Eqs. \eqref{eq:lim_Set}, \eqref{eq:lim_Tet}, \eqref{eq:lim_Seu} and \eqref{eq:lim_Teu}, that can be seen in Fig. \ref{fig:global_lom}. In addition, a possibility of different values of $\Delta m^2_{31}$, the summary of the limits on the NSI and on the standard neutrino oscillation parameters can be seen in Fig. \ref{fig:global_lom}.

\begin{acknowledgments}
 P.C.H and O.L.G.P. were thankful for the support of FAPESP funding Grant 2014/19164-6. O.L.G.P were thankful for the support of  FAEPEX funding grant 2391/2017 and 2541/2019, CNPq grant 306565/2019-6. M.E.C. is thankful for 140564/2018-7 funding from CNPQ and 88887.477371/2020-00 funding from CAPES. The authors are thankful to Zahra Tabrizi, Marcelo Guzzo and Yago Porto for useful discussions about the topic. This study was financed in part by the Coordenação de Aperfeiçoamento de Pessoal de Nível Superior - Brasil (CAPES) - Finance Code 001.
\end{acknowledgments}

\bibliographystyle{jhep}
\bibliography{nsidp}

\appendix

\section{Analytical expressions}
\label{sec:analytical}

After a straightforward calculation, here explicitly computed for the case $\alpha=\beta=e$, we find the $N^{\rm non-osc}$, $N^{\rm osc}$, and $N^{\rm CP}$ terms in each oscillation scale ($\Delta m^2_{21}$ and $\Delta m^2_{3i}$) and for different NSI couplings each time ($[\tilde{\epsilon}_X]_{e\mu}$ and $[\tilde{\epsilon}_X]_{e\tau}$). We will not consider mixing terms like $[\tilde{\epsilon}_X]_{e\tau}[\tilde{\epsilon}_X]_{e\mu}$, therefore our analysis is focused on the effects of the effective parameters. 

First, we present the non-oscillation terms, and the term will be the same for the scenario of  $[\tilde{\epsilon}_X]_{e\alpha}$) for $\alpha=e,\mu$, and they do not depend on the oscillation scale. They can appear, for example, in beta decays or reactor short baseline experiments:
\begin{equation}
        N^{\rm non-osc} = 1+2\left|[\tilde{\epsilon}_X]_{e\alpha}\right|^2d_{\rm XL}p_{\rm XL}+\left|[\tilde{\epsilon}_X]_{e\alpha}\right|^4d_{\rm XX}p_{\rm XX},
\end{equation}
where $\alpha$ can be $\mu$ or $\tau$. For a linear expansion we have $ N^{\rm non-osc}\to 1$, exactly as in the standard neutrino oscillations. For general $[\tilde{\epsilon}_X]_{e\alpha}\neq 0$, we have the so-called {\it zero distance effect}~\cite{Langacker:1988up,Antusch_2006,Fernandez_Martinez_2007,Kopp_2008,Rodejohann_2010,Ohlsson:2012kf,Guzzo:2013tca,Agarwalla:2014bsa,EscrihuelaFerrandiz:2016pun,Falkowski:2019kfn,Ellis:2020ehi} as discussed in section~(\ref{sec:reactors}).

Secondly, we present the amplitude of the atmospheric scale terms, that is, the terms that have the oscillation phase is proportional to  $\Delta m^2_{3i}$, where i=1,2.
In the {\it one mass dominance scenario}~\cite{DERUJULA198054,Barger:1980um,Fogli:1995uu}, that adequate for medium baseline reactors experiments for standard neutrino oscillation,   we can use 
$\Delta m^2_{32}=\Delta m^2_{31}+\Delta m^2_{21}\approx \Delta m^2_{31}$. In this limit we have that, 

\subsubsection*{Atmospheric scale: $\tilde{\epsilon}_{e\mu}$}
To be in the atmospheric scale it implies that $k=3 l=1,2$ or vice-versa. The results are
\begin{eqnarray}
   N^{\rm osc}_{\rm atm} =& s_{2\theta_{13}}^2 \left( 1+\frac{c_{13}^2}{s_{13}}\left|[\tilde{\epsilon}_X]_{e\mu}\right|^2d_{\rm XL}p_{\rm XL}\right)\label{eq:euatm}\\
   N^{\rm CP}_{\rm atm} =& 0&
\end{eqnarray}
where the first term is the standard neutrino oscillation and the others are the NSI terms.
\subsubsection*{Atmospheric scale: $\tilde{\epsilon}_{e\tau}$}

\begin{align}
   N^{\rm osc}_{\rm atm} =&
s_{2\theta_{13}}^2  \left(1 +
   \frac{c_{2\theta_{13}}}{s_{2\theta_{13}}}\Re{[\tilde{\epsilon}_X]_{e\tau}}(d_{\rm XL}+p_{\rm XL})-4\left(\Re{[\tilde{\epsilon}_X]_{e\tau}}\right)^2
   d_{\rm XL}p_{\rm XL} \right)
   +4 \left|[\tilde{\epsilon}_X]_{e\tau}\right|^2 d_{\rm XL}p_{\rm XL}
   \nonumber\\  
   &- s_{2\theta_{13}}^2\left|[\tilde{\epsilon}_X]_{e\tau}\right|^2 \left(
   d_{\rm XX} + p_{\rm XX}+c_{2\theta_{13}}^2\Re{[\tilde{\epsilon}_X]_{e\tau}}( d_{\rm XX}p_{\rm XL} + p_{\rm XX}d_{\rm XL})-\left|[\tilde{\epsilon}_X]_{e\tau}\right|^2 d_{\rm XX}p_{\rm XX}\right)
   \label{eq:etauatm}\\
   N^{\rm CP}_{\rm atm} =& +\Im{[\tilde{\epsilon}_X]_{e\tau}} s_{2\theta_{13}}
   \left( 
   (d_{\rm XL}-p_{\rm XL})+
   \left|[\tilde{\epsilon}_X]_{e\tau}\right|^2  s_{2\theta_{13}} (d_{\rm XX}p_{\rm XL}-d_{\rm XL}p_{\rm XX})
   \right) 
   \label{eq:atmCP}
\end{align}
As can be seen from Eq. \eqref{eq:euatm} and \eqref{eq:atmCP}, CP violation term at atmospheric scale appears only for the NSI with non-zero $e\tau$ parameter. We also expect that, if the NSI coupling is small, the $e\tau$ parameters are more important in the atmospheric scale than the $e\mu$. This comes from the fact that the $e\tau$~(eq.~\ref{eq:etauatm}) appears at linear order, opposite to $e\mu$~(eq.~\ref{eq:euatm}) that appears only at quadratic order.

\subsubsection*{Breaking the {\it one mass dominance scenario}}

Next, we will present the behavior of the atmospheric scale in the case we break {\it one mass dominance scenario}, that is, when we can consider $\Delta m^2_{32}\neq \Delta m^2_{31}$.  When we dont assume the 
 {\it one mass dominance scenario}, the $N^{\rm osc}_{\rm 3i}$ are different for $i=1,2$ and are proprortional 
 to the coefficients, $  N^{\rm osc}_{\rm atm}$ and  $  N^{\rm osc}_{\rm CP}$ as 
\begin{align}
N^{\rm osc}_{\rm 32} = N^{\rm osc}_{\rm atm}s^2_{12},& & N^{\rm osc}_{\rm 31} = N^{\rm osc}_{\rm atm}c^2_{12},\\ N^{\rm CP}_{\rm 32} = N^{\rm CP}_{\rm atm} s^2_{12},& &N^{\rm CP}_{\rm 31} = N^{\rm CP}_{\rm atm}c^2_{12}. 
\end{align}
In contrast, for the $e\mu$ parameter, we have more complex changes:
\begin{align}
   N^{\rm osc}_{\rm 32} =&s_{2\theta_{13}}^2 s^2_{12}{+{\Re }\{[\tilde{\epsilon}_X]_{e\mu}\}(p_{\rm XL}+d_{\rm XL})s_{2\theta_{12}}c_{13}s^2_{13}} +4\left|[\tilde{\epsilon}_X]_{e\mu}\right|^2d_{\rm XL}p_{\rm XL}s_{13}c^2_{12}\label{xeq1a} \\
   N^{\rm CP}_{\rm 32} =& -\Im \{[\tilde{\epsilon}_X]_{e\mu}\}(p_{\rm XL}-d_{\rm XL}) s_{2\theta_12}c_{13}s^4_{13},
   \label{xeq1b}
\end{align}
\begin{align}
   N^{\rm osc}_{\rm 31} =& s_{2\theta_{13}}^2c^2_{12}{-\Re \{[\tilde{\epsilon}_X]_{e\mu}\}(p_{\rm XL}+d_{\rm XL})s_{2\theta_{12}}c_{13}s^2_{13} +4\left|[\tilde{\epsilon}_X]_{e\mu}\right|^2d_{\rm XL}p_{\rm XL} s_{13}s^2_{12}}    \label{xeq2a}\\
   N^{\rm CP}_{\rm 31} =& +\Im \{[\tilde{\epsilon}_X]_{e\mu}\}(p_{\rm XL}-d_{\rm XL})s_{2\theta_{12}}c_{13}s^4_{13}
      \label{xeq2b}
\end{align}
The more remarkable of this results is the NSI contribution have equal coefficients but with opposite signs for the $\Delta m_{31}^2$ and for  $\Delta m_{32}^2$ and results into beating effect, and these combination is proportional to $\Delta m_{21}^2$.

\subsubsection*{Solar scale}

Finally, we present the rates for the solar scale, that is, terms that are proportional to $\Delta m^2_{21}$: 

\subsubsection*{Solar scale: $\tilde{\epsilon}_{e\mu}$}

\begin{align}
    N^{\rm osc}_{\rm Sun} =& c_{13}^4 s_{2\theta_{12}}^2
    -\Re{[\tilde{\epsilon}_X]_{e\mu}} c_{13}^3 s_{4\theta_{12}}(d_{\rm XL}+p_{\rm XL})\nonumber
    \\&+\left(\left|[\tilde{\epsilon}_X]_{e\mu}\right|^2+\Re{[\tilde{\epsilon}_X]_{e\mu}}^2 s_{2\theta_{12}}^2\right) 4c_{13}^2 d_{\rm XL}p_{\rm XL}\nonumber
    -\left|[\tilde{\epsilon}_X]_{e\mu}\right|^2 s_{2\theta_{12}}^2( d_{\rm XX} + p_{\rm XX})\nonumber
    \\&-\left|[\tilde{\epsilon}_X]_{e\mu}\right|^2\Re{[\tilde{\epsilon}_X]_{e\mu}} c_{13}s_{4\theta_{12}}^2( d_{\rm XX}p_{\rm XL} + p_{\rm XX}d_{\rm XL})+\left|[\tilde{\epsilon}_X]_{e\mu}\right|^4 s_{2\theta_{12}}^2  d_{\rm XX}p_{\rm XX}
    \label{eq:eusolar}
\end{align}

\begin{align}
   N^{\rm CP}_{\rm Sun} = +\left[c_{13}^2(d_{\rm XL}-p_{\rm XL})+\left|[\tilde{\epsilon}_X]_{e\mu}\right|^2(d_{\rm XX}p_{\rm XL}-d_{
   \rm XL}p_{\rm XX})\right]\Im{[\tilde{\epsilon}_X]_{e\mu}}c_{13}{s_{2\theta_{12}}}
  \label{eq:solarCP}
\end{align}
\subsubsection*{Solar scale: $\tilde{\epsilon}_{e\tau}$}


\begin{align}
    N^{\rm osc}_{\rm Sun} =& c_{13}^4 s_{2\theta_{12}}^2-2s_{13}c_{13}^3\Re{[\tilde{\epsilon}_X]_{e\tau}}(d_{\rm XL}+p_{\rm XL}) s_{2\theta_{12}}^2\nonumber\\&+s_{2\theta_{13}}^2s_{2\theta_{12}}^2\Re{[\tilde{\epsilon}_X]_{e\tau}}^2 d_{\rm XL}p_{\rm XL}\nonumber+\frac{1}{4}\left|[\tilde{\epsilon}_X]_{e\tau}\right|^2 s_{2\theta_{12}}^2 4 s_{13}^2 c_{13}^2( d_{\rm XX} + p_{\rm XX})\nonumber\\&-c_{13}s_{13}^3\left|[\tilde{\epsilon}_X]_{e\tau}\right|^2s_{4\theta_{12}}^2\Re{[\tilde{\epsilon}_X]_{e\tau}}( d_{\rm XX}p_{\rm XL} + p_{\rm XX}d_{\rm XL})\nonumber\\&+s_{13}^4\left|[\tilde{\epsilon}_X]_{e\tau}\right|^4d_{\rm XX}p_{\rm XX}s_{2\theta_{12}}^2 
       \label{eq6}\\
       N^{\rm CP}_{\rm Sun} =& 0.
           \label{eq7}
\end{align}
\section{Rephasing invariant with non-standard neutrino interaction}
\label{invariants}
In standard neutrino oscillation induced by PMNS mixing matrix there are invariant by phase rephasing~\cite{ 2014JPhCS.485a2058P} called Jarskolg invariant~\cite{PhysRevLett.55.1039}.
We follow  Ref.~(\cite{SARKAR200728}) to construct the invariant by global phase rephasing  for our BSM Lagrangian. If we made global rephasing the particle fields 
\begin{equation}
\begin{aligned}
\nu_i \to e^{i\delta_i}\nu_i & \quad l_{\alpha} \to e^{i\eta_{\alpha}} l_{\alpha}
\nonumber \\
u\to e^{i\delta_u}u & \quad
d\to e^{i\delta_d}d \nonumber 
\end{aligned}
\label{seseq7}
\end{equation}
then the Lagrangian  given in Eq.~(\ref{eq:EFT_lweft}) it will invariant if 
\begin{equation}
\begin{aligned}
V_{\rm ud}\to V_{\rm ud} e^{i(\delta_u-\delta_d)} &\quad
U_{\alpha j}\to U_{\alpha j} e^{i(\eta_{\alpha}-\delta_j)}\nonumber \\
(\epsilon_S)_{\alpha\beta} \to (\epsilon_S)_{\alpha\beta} e^{+i(\eta_{\alpha}-\eta_{\beta})} &\quad
(\epsilon_T)_{\alpha\beta} \to (\epsilon_T)_{\alpha\beta} e^{+i(\eta_{\alpha}-\eta_{\beta})}
\end{aligned}
\label{seseq8}
\end{equation}
Now, with these rules we can apply to the with NSI rate~(\ref{eq:rates_as_prob}) and we have found that the each term of the rate is is invariant by these global re-phasing.

\section{Solar+KamLand results}
\label{sec:solar_analysis}

\begin{figure}[ht!]
    \centering
    \includegraphics[scale=0.85]{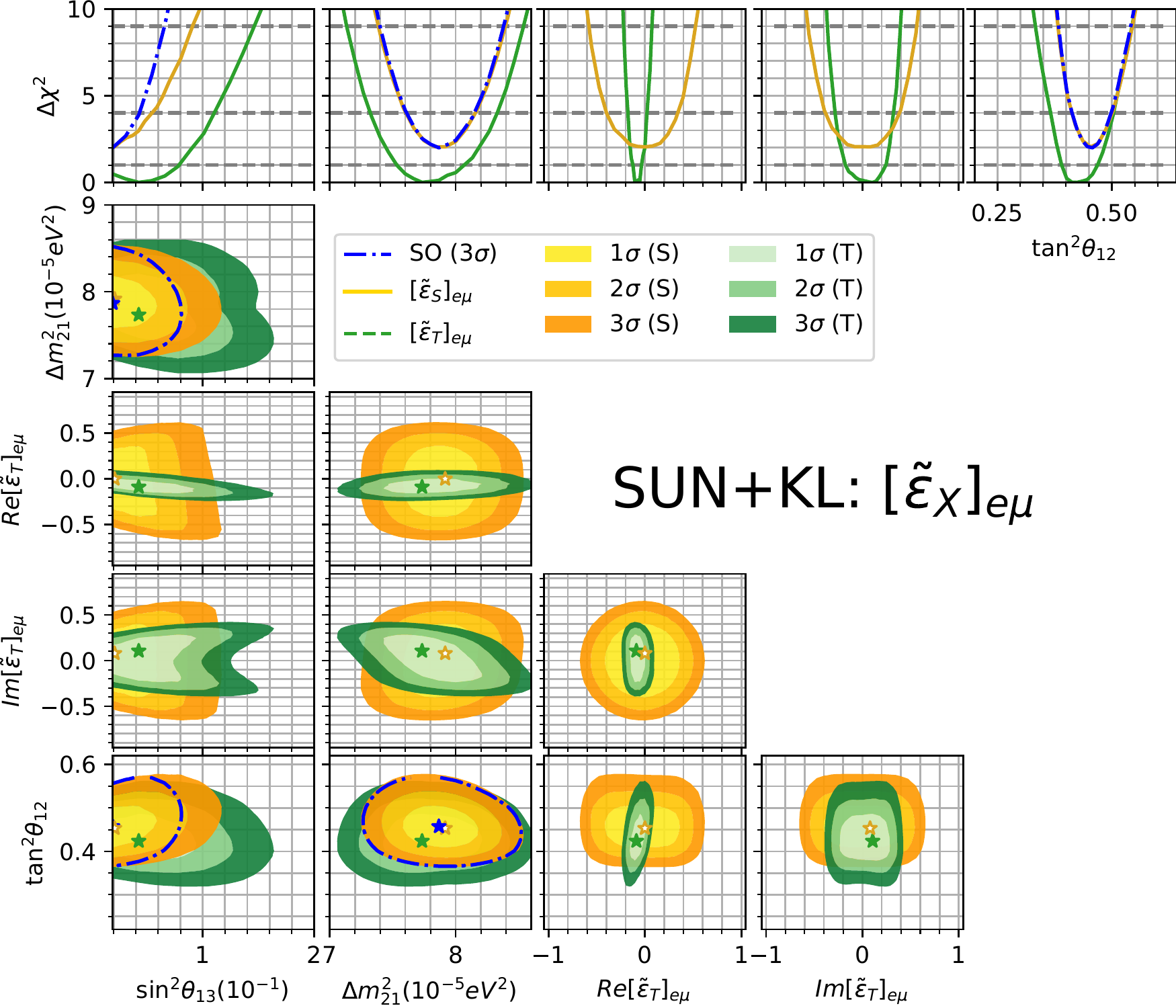}
    \caption{The $\Delta \chi^2$ of KamLand combined with solar experiments in the presence of $[\tilde{\epsilon}_X]_{e\tau}$, the contour levels at $1\sigma$, $2\sigma$ and $3\sigma$ and the 1D $\chi^2$ curves. In green (yellow) curve  is the NSI tensor (scalar) case and in dash-dotted blue curve is $\chi^2_{\rm SO}-\text{min}(\chi^2_{\rm NSI}).$}
    \label{fig:solar-eu}
\end{figure}

Here we present the Solar+KamLand analysis using the $\chi_{\rm KL}^2$ and the $\chi_{\rm SUN}^2$. The results of the 1D and 2D $\chi^2$ functions are shown in fig. \ref{fig:solar-eu}. In the first line we show the $\chi^2$ functions with the difference of the $(\chi^2([\tilde{\epsilon}_T]_{e\mu}))_{\rm min}$. In green, we show the tensor interactions, and in yellow, the scalar interactions. For the scalar interactions, there are no substantial differences with the standard neutrino oscillation. However, in the tensor case, there is an slight improvement of around $1.3\sigma$ in the $\Delta \chi^2$ compared to the standard neutrino oscillation model. 

In the Fig. \ref{fig:solar-eu} we show the 2D contour which shows the correlations between all the parameters considered in the analysis. For each pair of variables we minimize of the complementary parameter space. Let's discuss separately each interaction, scalar and tensor:

\begin{itemize}
    \item For the $[\tilde{\epsilon}_S]_{e\mu}$ parameter, there was no improvement in the fit compared to the standard neutrino oscillation. The only change is a small distortion in the $\theta_{13}$. The limits on the NSI parameters from solar+KamLand experiment are:
    
    \begin{equation}
    \begin{aligned}
        \Re [\tilde{\epsilon}_S]_{e\mu} = 0.00_{ -0.30}^{+0.27},  & & \Im [\tilde{\epsilon}_S]_{e\mu} = +0.01_{-0.29}^{+0.29}. 
    \end{aligned}
    \label{eq:lim_Seu_sol}
    \end{equation}
    
    \item For the $[\tilde{\epsilon}_T]_{e\mu}$ parameter, there was a $1.3\sigma$ improvement in the fit compared to the standard neutrino oscillation. The improvement in the fit comes from a non-zero $\Re[\tilde{\epsilon}_T]_{e\mu}$. The limits on the NSI parameters from solar+Kamland experiment are:
    
    \begin{equation}
    \begin{aligned}
        \Re [\tilde{\epsilon}_T]_{e\mu} = -0.09_{ -0.03}^{+0.05},  & & \Im [\tilde{\epsilon}_T]_{e\mu} = +0.10_{-0.09}^{+0.14}. 
    \end{aligned}
    \label{eq:lim_Teu_sol}
    \end{equation}
    
\end{itemize}

\begin{figure}[ht!]
    \centering
    \includegraphics[scale=0.85]{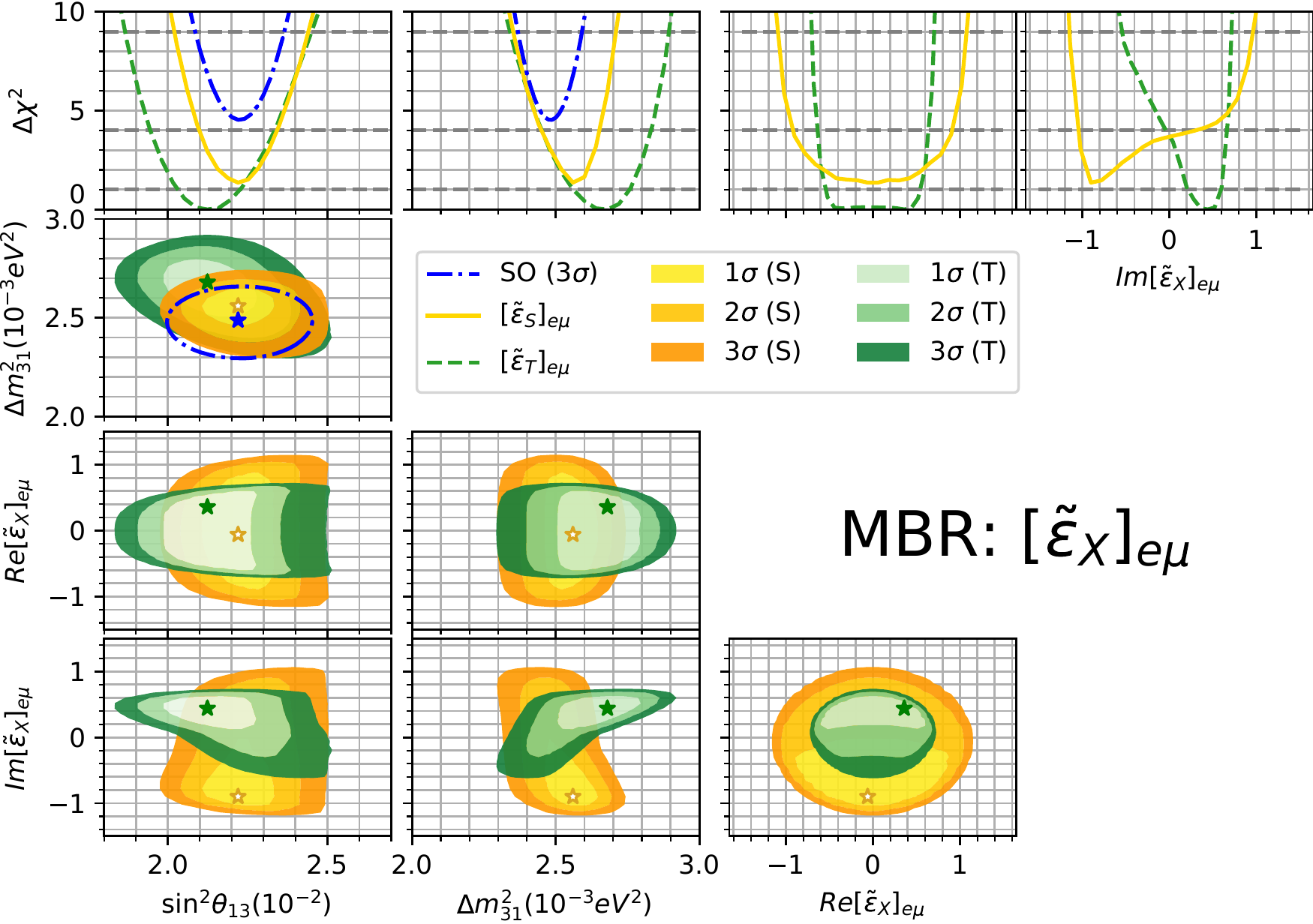}
    \caption{The same as in fig. \ref{fig:MBL-et} but for $[\tilde{\epsilon}_X]_{e\mu}$ interactions.}
    \label{fig:MBL-eu}
\end{figure}

\section{$[\tilde{\epsilon}_X]_{e\mu}$ effects on Medium baseline reactor}
\label{sec:MBR_eu_analysis}
In the $[\tilde{\epsilon}_X]_{e\mu}$ MBR analysis, we found an improvement in the statistics when comparing with the standard neutrino oscillation $1.8\sigma$ in the scalar interaction case and $2.1$ in the scalar case. In figure \ref{fig:MBL-eu}, we show a grid of graphs where in the first line we show the $\Delta \chi^2$ functions where in blue is the standard neutrino oscillation, in yellow scalar interactions and in green tensor interactions. The improvement in the both fits come mainly from the imaginary part of the $[\tilde{\epsilon}_X]_{e\mu}$ that cause a non-zero CP violation effect, Eq. \eqref{eq:CP_new}, that affects the MBR experiments. As this is an effect that comes from the solar sector of the oscillation probability, the $\theta_{13}$ and $\Delta m^2_{31}$ does not change in the same size as in the $[\tilde{\epsilon}_X]_{e\tau}$ analysis, see the $\sin^2\theta_{13}\times \Delta m^2_{31}$ panel of Fig.~\ref{fig:MBL-eu}. In Fig.~\ref{fig:MBL-eu} we show the $\chi^2$ contours for the 2D analysis of MBR reactors, for each panel we marginalize over the complementary set of parameter. Let's discuss in details each interactions case, scalar and tensor:

\begin{itemize}
    \item The scalar $[\tilde{\epsilon}_S]_{e\mu}$ interactions, leads to an improvement of $1.8\sigma$ in the analysis. There is no correlation between the variables and improvement in the fit come alone from the $\Im[\tilde{\epsilon}_S]_{e\mu}$, hence, this is a pure CP violation effect. The values of NSI are: 

    \begin{equation}
    \begin{aligned}
        \Re [\tilde{\epsilon}_S]_{e\mu} = 0.00_{ -0.67}^{+0.67},  & & \Im [\tilde{\epsilon}_S]_{e\mu} = -0.87_{-0.08}^{+0.37}. 
    \end{aligned}
    \label{eq:lim_Seu_MBR}
    \end{equation}

    \item The scalar $[\tilde{\epsilon}_T]_{e\mu}$ interactions, leads to an improvement of $2.1\sigma$ in the analysis. There is a small correlation  between the $\Im[\tilde{\epsilon}_T]_{e\mu}$ and the standard neutrino oscillation variables $\sin^2\theta_{13}$ (negatively) and $\Delta m^2_{31}$ (positively). As in the scalar case, the improvement in the fit comes mainly from the $\Im[\tilde{\epsilon}_T]_{e\mu}$ with a small preference for non-zero NSI. The values of NSI are: 

    \begin{equation}
    \begin{aligned}
        \Re [\tilde{\epsilon}_T]_{e\mu} = 0.37_{ -0.91}^{+0.16},  & & \Im [\tilde{\epsilon}_T]_{e\mu} = +0.44_{-0.20}^{+0.17}. 
    \end{aligned}
    \label{eq:lim_Teu_MBR}
    \end{equation}

\end{itemize}

\end{document}